%% file: beijing_ppe.tex
\documentstyle[epsfig,12pt]{article}

\input{ graphics_sizes }
\input{ eps_graphics }
\input{ graftwos_macro }
\newlength{\peteheight}
\multiply \peteheight by 5
\divide \peteheight by 4

\catcode`\@=11
\long\def\@makefntext#1{
\protect\noindent \hbox to 3.2pt {\hskip-.9pt
$^{{\ninerm\@thefnmark}}$\hfil}#1\hfill}                %CAN BE USED

\def\@makefnmark{\hbox to 0pt{$^{\@thefnmark}$\hss}}  %ORIGINAL
        
\def\ps@myheadings{\let\@mkboth\@gobbletwo
\def\@oddhead{\hbox{}
\rightmark\hfil\ninerm\thepage}
\def\@oddfoot{}\def\@evenhead{\ninerm\thepage\hfil
\leftmark\hbox{}}\def\@evenfoot{}
\def\sectionmark##1{}\def\subsectionmark##1{}}

\renewcommand{\thefootnote}{\fnsymbol{footnote}}

\newcounter{sectionc}\newcounter{subsectionc}\newcounter{subsubsectionc}
\renewcommand{\section}[1] {\vspace*{0.6cm}\addtocounter{sectionc}{1}
\setcounter{subsectionc}{0}\setcounter{subsubsectionc}{0}\noindent
        {\normalsize\bf\thesectionc. #1}\par\vspace*{0.4cm}}
\renewcommand{\subsection}[1] {\vspace*{0.6cm}\addtocounter{subsectionc}{1}
        \setcounter{subsubsectionc}{0}\noindent
        {\normalsize\it\thesectionc.\thesubsectionc. #1}\par\vspace*{0.4cm}}
\renewcommand{\subsubsection}[1]
{\vspace*{0.6cm}\addtocounter{subsubsectionc}{1}
        \noindent {\normalsize\rm\thesectionc.\thesubsectionc.\thesubsubsectionc.
        #1}\par\vspace*{0.4cm}}

\newcounter{appendixc}
\newcounter{subappendixc}[appendixc]
\newcounter{subsubappendixc}[subappendixc]

\renewcommand{\appendix}[1] {\vspace*{0.6cm}
        \refstepcounter{appendixc}
        \setcounter{figure}{0}
        \setcounter{table}{0}
        \setcounter{equation}{0}
        \renewcommand{\thefigure}{\Alph{appendixc}.\arabic{figure}}
        \renewcommand{\thetable}{\Alph{appendixc}.\arabic{table}}
        \renewcommand{\theappendixc}{\Alph{appendixc}}
        \renewcommand{\theequation}{\Alph{appendixc}.\arabic{equation}}
        \noindent{\bf Appendix \theappendixc #1}\par\vspace*{0.4cm}}

\def\abstracts#1{{
        \centering{\begin{minipage}{12.2truecm}\footnotesize\baselineskip=12pt\noindent
        \centerline{\footnotesize ABSTRACT}\vspace*{0.3cm}
        \parindent=0pt #1
        \end{minipage}}\par}}

\renewenvironment{thebibliography}[1]
        {\begin{list}{\arabic{enumi}.}
        {\usecounter{enumi}\setlength{\parsep}{0pt}
\setlength{\leftmargin 1.25cm}{\rightmargin 0pt}
         \setlength{\itemsep}{0pt} \settowidth
        {\labelwidth}{#1.}\sloppy}}{\end{list}}

\newcounter{itemlistc}
\newcounter{romanlistc}
\newcounter{alphlistc}
\newcounter{arabiclistc}

\newcommand{\fcaption}[1]{
        \refstepcounter{figure}
        \setbox\@tempboxa = \hbox{\footnotesize Fig.~\thefigure. #1}
        \ifdim \wd\@tempboxa > 6in
           {\begin{center}
        \parbox{6in}{\footnotesize\baselineskip=12pt Fig.~\thefigure. #1}
            \end{center}}
        \else
             {\begin{center}
             {\footnotesize Fig.~\thefigure. #1}
              \end{center}}
        \fi}

\newcommand{\tcaption}[1]{
        \refstepcounter{table}
        \setbox\@tempboxa = \hbox{\footnotesize Table~\thetable. #1}
        \ifdim \wd\@tempboxa > 6in
           {\begin{center}
        \parbox{6in}{\footnotesize\baselineskip=12pt Table~\thetable. #1}
            \end{center}}
        \else
             {\begin{center}
             {\footnotesize Table~\thetable. #1}
              \end{center}}
        \fi}

\def\@citex[#1]#2{\if@filesw\immediate\write\@auxout
        {\string\citation{#2}}\fi
\def\@citea{}\@cite{\@for\@citeb:=#2\do
        {\@citea\def\@citea{,}\@ifundefined
        {b@\@citeb}{{\bf ?}\@warning
        {Citation `\@citeb' on page \thepage \space undefined}}
        {\csname b@\@citeb\endcsname}}}{#1}}

\newif\if@cghi
\def\cite{\@cghitrue\@ifnextchar [{\@tempswatrue
        \@citex}{\@tempswafalse\@citex[]}}
\def\citelow{\@cghifalse\@ifnextchar [{\@tempswatrue
        \@citex}{\@tempswafalse\@citex[]}}
\def\@cite#1#2{{$\null^{#1}$\if@tempswa\typeout
        {IJCGA warning: optional citation argument
        ignored: `#2'} \fi}}

 1
 1
 1

\font\ninerm=cmr9

\begin{document}
\centerline{\Large UNIVERSITY OF WISCONSIN -- MADISON}
\vspace*{.3in}
\begin{flushright}
WISC-EX-96--343\\
1 February 1996
\end{flushright}

\vspace*{0.6in}
\baselineskip=15pt
{\mathversion{bold}
{ \centerline{\Large\bf Recent Results on $B$\/ Meson Oscillations}}
 \mathversion{normal}}
\vspace{.6in}

%\vfill
\baselineskip=13pt
\centerline{\normalsize SAU LAN WU}
\baselineskip=13pt
\centerline{\normalsize\it Physics Department, University of Wisconsin--
Madison}
\baselineskip=12pt
\centerline{\normalsize\it Madison, WI, 53706, USA}

\vspace*{.5in}

%\vfill
\baselineskip=13pt
\centerline{\bf Abstract}
\vspace*{0.3in}
This paper presents recent time-dependent measurements of
neutral $B$\/ meson oscillations.  Similar to the $K^{0}$--$\bar{K}^{0}$\/
system, there are two such systems involving the $b$\/ quark:
$B_{d}^{0}$--$\bar{B}_{d}^{0}$ and $B_{\mbox{\scriptsize s}}^{0}$--$\bar{B}_{\mbox{\scriptsize s}}^{0}$\/.
Thus the physical states are respectively $K_{S}$\/ and $K_{L}$\/,
$(B_{d})_{S}$\/ and $(B_{d})_{L}$\/, and
$(B_{\mbox{\scriptsize s}})_{S}$\/ and $(B_{\mbox{\scriptsize s}})_{L}$\/.  The oscillation between each
pair of states can be used to determine their mass difference.
The present world average for the $(B_{d})_{S}$--$(B_{d})_{L}$
mass difference is $\Delta m_{d}$\/~=~0.457$\pm$\/0.019 ps$^{-1}$
(or (3.01$\pm$\/0.13)$\times 10^{-4} \: eV$\/).  Using
$f_{B_{\mbox{\scriptsize s}}}$\/ = 12\% (the fraction of $B_{\mbox{\scriptsize s}}$ produced in $b$\/ events),
the current lower limit on the corresponding $\Delta m_{\mbox{\scriptsize s}}$\/
is 6.1 ps$^{-1}$ (or 4.0$\times 10^{-3} \: eV$\/).

%    *********** after abstract

\vfill
\vspace{.2in}
\centerline{(Invited talk at the XVII International Symposium on}
\centerline{Lepton-Photon Interactions, Beijing, China, August 1995.)}

\clearpage
\pagenumbering{arabic}
\pagestyle{plain}
\baselineskip=15pt
\centerline{\normalsize\bf Recent Results on ${\bf B}$\/ Meson Oscillations}

%\vfill
\vspace*{0.3cm}
\baselineskip=13pt
\centerline{\footnotesize SAU LAN WU}
\baselineskip=13pt
\centerline{\footnotesize\it Physics Department, University of Wisconsin--
Madison}
\baselineskip=12pt
\centerline{\footnotesize\it Madison, WI, 53706, USA}
\centerline{\footnotesize E-mail: WUS@cernvm.cern.ch}
\vspace*{0.9cm}
\baselineskip=13pt
\abstracts{This paper presents recent time-dependent measurements of
neutral $B$\/ meson oscillations.  Similar to the $K^{0}$--$\bar{K}^{0}$\/
system, there are two such systems involving the $b$\/ quark:
$B_{d}^{0}$--$\bar{B}_{d}^{0}$ and $B_{\mbox{\scriptsize s}}^{0}$--$\bar{B}_{\mbox{\scriptsize s}}^{0}$\/.
Thus the physical states are respectively $K_{S}$\/ and $K_{L}$\/,
$(B_{d})_{S}$\/ and $(B_{d})_{L}$\/, and
$(B_{\mbox{\scriptsize s}})_{S}$\/ and $(B_{\mbox{\scriptsize s}})_{L}$\/.  The oscillation between each
pair of states can be used to determine their mass difference.
The present world average for the $(B_{d})_{S}$--$(B_{d})_{L}$
mass difference is $\Delta m_{d}$\/~=~0.457$\pm$\/0.019 ps$^{-1}$
(or (3.01$\pm$\/0.13)$\times 10^{-4} \: eV$\/).  Using
$f_{B_{\mbox{\scriptsize s}}}$\/ = 12\% (the fraction of $B_{\mbox{\scriptsize s}}$ produced in $b$\/ events),
the current lower limit on the corresponding $\Delta m_{\mbox{\scriptsize s}}$\/
is 6.1 ps$^{-1}$ (or 4.0$\times 10^{-3} \: eV$\/).}

%\vspace*{0.6cm}
\normalsize\baselineskip=15pt
\setcounter{footnote}{0}
\renewcommand{\thefootnote}{\alph{footnote}}
\section{Introduction}
Since there are three known quarks of charge $-\frac{1}{3}$, namely $d$\/,
$s$\/ and $b$\/, there are three similar neutral particle--antiparticle
systems:
\begin{displaymath}
\begin{array}{cccc}
K^{0} (\bar{s}d)$--$\bar{K^{0}} (s\bar{d}), &
B_{d} (\bar{b}d)$--$\bar{B_{d}} (b\bar{d}), &
{\rm and } &
B_{\mbox{\scriptsize s}} (\bar{b}s)$--$\bar{B_{\mbox{\scriptsize s}}} (b\bar{s}).
\end{array}
\end{displaymath}
Of these three, the $K^{0}$\/--$\bar{K^{0}}$\/ system is best measured
experimentally\cite{OLD:kaon,RF:PDG} and understood theoretically\cite{TH:WuYang}.
However, the theoretical
analysis applies equally well to all three cases.

As the topic here is the $B\bar{B}$ mixing, let $B$ and
$\bar{B}$ denote the flavor state in all three cases, i.e.,
\begin{displaymath}
\begin{array}{ccccccc}
    &   B & = & K^{0}       & B_{d}       & {\it or }& B_{\mbox{\scriptsize s}} \\
{\it and } &\bar{B} & = & \bar{K}^{0} & \bar{B}_{d} & {\it or } & \bar{B}_{\mbox{\scriptsize s}},
\end{array}
\end{displaymath}
while the corresponding weak eigenstates are
\begin{displaymath}
\begin{array}{ccccccc}
           &       B_{S}  & = & K_{S}, & (B_{d})_{S} 
 & {\it or } & (B_{\mbox{\scriptsize s}})_{S} \\
{\it and } &       B_{L}  & = & K_{L}, & (B_{d})_{L}
 & {\it or } & (B_{\mbox{\scriptsize s}})_{L}.
\end{array}
\end{displaymath}
Because of CP non-conservation, $B_{S}$ and $B_{L}$, which are not
orthogonal, can most generally be related to $B$ and $\bar{B}$ by
\begin{displaymath}
\begin{array}{cccc}
          & B_{S} & = & \left( \left| p \right|^{2} +
                        \left| q \right|^{2} \right)^{-\frac{1}{2}}
                        \left( p B + q \bar{B} \right) \\
{\it and} & B_{L} & = & \left( \left| p \right|^{2} +
                        \left| q \right|^{2} \right)^{-\frac{1}{2}}
                        \left( p B - q \bar{B} \right).
\end{array}
\end{displaymath}

Let
\( \Gamma = \left( \Gamma_{S} + \Gamma_{L} \right) / 2 \)
and
\( m = \left( m_{S} + m_{L} \right) / 2 \)
be the average width and mass of $B_{S}$
and $B_{L}$, while $\Delta \Gamma$ and $\Delta m$ are the differences
\begin{displaymath}
\Delta \Gamma = \Gamma_{S} - \Gamma_{L} > 0
\end{displaymath}
and
\begin{displaymath}
\Delta m = \left| m_{S} - m_{L} \right| .
\end{displaymath}
Let ${\cal P}_{B,u}(t)$ and ${\cal P}_{\bar{B},u}(t)$ be the probability
distributions for a meson which is created as $B$ (or $\bar{B}$) to
decay as a $B$ (or $\bar{B}$) after a proper time $t$, and
${\cal P}_{B,m}(t)$ and ${\cal P}_{\bar{B},m}(t)$ be those for a meson
created as $B$ (or $\bar{B}$) to decay as a $\bar{B}$ (or $B$), where
the subscripts $u$ and $m$ stand for {\em unmixed} and {\em mixed}
respectively.  These four quantities are given by
\begin{eqnarray}
 {\cal P}_{B,u}\left(t\right) & = &
   \frac{\left| p \right|^{2}}
        {\Gamma
         \left[
          \frac{ \left| p \right|^{2} + \left| q \right|^{2}}
               {\Gamma^{2} - \left( \Delta \Gamma / 2 \right)^{2}}
        +
          \frac{ \left| p \right|^{2} - \left| q \right|^{2}}
               {\Gamma^{2} + \left( \Delta m \right)^{2}}
         \right]}
   e^{-\Gamma t}
   \left[ \cosh{ \frac{\Delta \Gamma}{2}\, t} + \cos{\Delta m \, t} \right] \nonumber \\
{\cal P}_{B,m}(t) & = &
   \frac{\left| q \right|^{2}}
        {\Gamma
         \left[
          \frac{ \left| p \right|^{2} + \left| q \right|^{2}}
               {\Gamma^{2} - \left( \Delta \Gamma / 2 \right)^{2}}
        +
          \frac{ \left| p \right|^{2} - \left| q \right|^{2}}
               {\Gamma^{2} + \left( \Delta m \right)^{2}}
         \right]}
   e^{-\Gamma t}
   \left[ \cosh{\frac{\Delta \Gamma}{2}\, t} - \cos{\Delta m \, t} \right] \nonumber \\
{\cal P}_{\bar{B},u}(t) & = &
   \frac{\left| q \right|^{2}}
        {\Gamma
         \left[
          \frac{ \left| p \right|^{2} + \left| q \right|^{2}}
               {\Gamma^{2} - \left( \Delta \Gamma / 2 \right)^{2}}
        -
          \frac{ \left| p \right|^{2} - \left| q \right|^{2}}
               {\Gamma^{2} + \left( \Delta m \right)^{2}}
         \right]}
   e^{-\Gamma t}
   \left[ \cosh{\frac{\Delta \Gamma}{2}\, t} + \cos{\Delta m \, t} \right] \nonumber \\
{\cal P}_{\bar{B},m}(t) & = &
   \frac{\left| p \right|^{2}}
        {\Gamma
         \left[
          \frac{ \left| p \right|^{2} + \left| q \right|^{2}}
               {\Gamma^{2} - \left(  \Delta \Gamma / 2 \right)^{2}}
        -
          \frac{ \left| p \right|^{2} - \left| q \right|^{2}}
               {\Gamma^{2} + \left( \Delta m \right)^{2}}
         \right]}
   e^{-\Gamma t}
   \left[ \cosh{\frac{\Delta \Gamma}{2}\, t} - \cos{\Delta m \, t} \right].
   \label{eq:propt}
\end{eqnarray}

Although these three systems, $K^{0}$--$\bar{K}^{0}$, $B_{d}$--$\bar{B}_{d}$,
and $B_{\mbox{\scriptsize s}}$--$\bar{B}_{\mbox{\scriptsize s}}$ can be described by the same set of equations,
the different decay modes lead to significant differences in the behavior
of these systems.  For $K^{0}$--$\bar{K}^{0}$,
the $\pi \pi$ mode dominates with the immediate consequence that
$\Gamma_{S} \gg \Gamma_{L}$; in fact,
\begin{displaymath}
\Gamma_{S}/\Gamma_{L} \sim 580.
\end{displaymath}
In contrast, both $B_{d}$ and $B_{\mbox{\scriptsize s}}$ have many important decay modes.
Indeed, for both cases the width difference $\Delta \Gamma$ comes from
decay modes that are available to both $B$ and $\bar{B}$.  Using the
many measured branching ratios for $B_{d}$, a generous estimate gives
\(\left(\Delta \Gamma / \Gamma \right)_{d} < 5 \%\), perhaps much less.
Since there is very little experimental information about the decay modes
of $B_{\mbox{\scriptsize s}}$\/, no such firm statement can be made about the ratio
\(\left(\Delta \Gamma / \Gamma \right)_{\mbox{\scriptsize s}}\), but it is also believed
to be small.  In this talk, the width difference $\Delta \Gamma$ will
be neglected for both the
$B_{d}$\/--$\bar{B}_{d}$\/ and the $B_{\mbox{\scriptsize s}}$--$\bar{B}_{\mbox{\scriptsize s}}$\/
systems.

From Eq.~(\ref{eq:propt}), if the effect of CP non-conservation is
neglected, then the formulas for the four probabilities simplify to
\begin{equation}
\begin{array}{ccccccc}
  {\cal P}_{u}         \left( t \right) & = &
  {\cal P}_{B,u}       \left( t \right) & = &
  {\cal P}_{\bar{B},u} \left( t \right) & = &
\frac{\Gamma^{2} - \left( \Delta \Gamma / 2 \right)^{2}}
     {2 \Gamma}
e^{- \Gamma t}
\left[
      \cosh{\frac{\Delta \Gamma}{2} \, t} +
      \cos{ \Delta m      \, t}
\right]  \nonumber \\
  {\cal P}_{m}         \left( t \right) & = &
  {\cal P}_{B,m}       \left( t \right) & = &
  {\cal P}_{\bar{B},m} \left( t \right) & = &
\frac{\Gamma^{2} - \left( \Delta \Gamma / 2 \right)^{2}}
     {2 \Gamma}
e^{- \Gamma t}
\left[
      \cosh{\frac{\Delta \Gamma}{2} \, t} -
      \cos{ \Delta m      \, t}
\right]. \label{eq:proptsim1}
\end{array}
\end{equation}

If, furthermore, $\Delta \Gamma$ is neglected, the preceeding
equations can be written as
\begin{eqnarray}
{\cal P}_{u} \left( t \right) & = &
\frac{\Gamma}
     {2}
e^{- \Gamma t}
\left[
      1 +
      \cos{ \Delta m      \, t}
\right] \nonumber \\
{\cal P}_{m} \left( t \right) & = &
\frac{\Gamma}
     {2}
e^{- \Gamma t}
\left[
      1 -
      \cos{ \Delta m      \, t}
\right]. \label{eq:proptsim2}
\end{eqnarray}

For the $K^{0}$--$\bar{K}^{0}$\/ system, the mass difference $\Delta m$\/
was measured a long time ago\cite{RF:PDG} to be
\( 3.51 \times 10^{-6} eV \)\/.  The purpose of this talk is to present
and discuss recent experimental results on $\Delta m_{d}$\/ and
$\Delta m_{\mbox{\scriptsize s}}$\/ for the $B_{d}$--$\bar{B}_{d}$\/ and
$B_{\mbox{\scriptsize s}}$--$\bar{B}_{\mbox{\scriptsize s}}$\/ systems.  The most sensitive measurements
of $\Delta m_{d}$\/ and $\Delta m_{\mbox{\scriptsize s}}$\/ are obtained through
{\em time-dependent}\/ measurements, which investigate ${\cal P}_{u}$\/
and ${\cal P}_{m}$\/ directly, and this talk will
consider only time-dependent measurements of these quantities.  Since
neither CP non-conservation nor the width differences $\Delta \Gamma$\/
have been observed in these systems, the analyses have been carried
out on the basis of Eq.~(\ref{eq:proptsim2}) rather than the more
accurate Eqs.~(\ref{eq:propt}) or (\ref{eq:proptsim1}).

\section{Experimental Overview}
In order to perform a time-dependent measurement of $\Delta m_{d}$\/
or $\Delta m_{\mbox{\scriptsize s}}$\/, one must measure the proper time of the decay of
the $B$\/ meson, and determine its {\em production flavor}\/ (i.e.,
$B$\/ or $\bar{B}$\/ at production) and {\em decay flavor}\/ (i.e.,
$B$\/ or $\bar{B}$\/ at decay) in order to ascertain whether the $B$\/
meson is mixed or unmixed.  The value of $\Delta m$\/ is
then found from the
fraction of events identified (or {\em tagged}\/) as mixed or unmixed
as a function of the measured proper time.
\begin{figure}[t]
\begin{center}
\epsfig{file=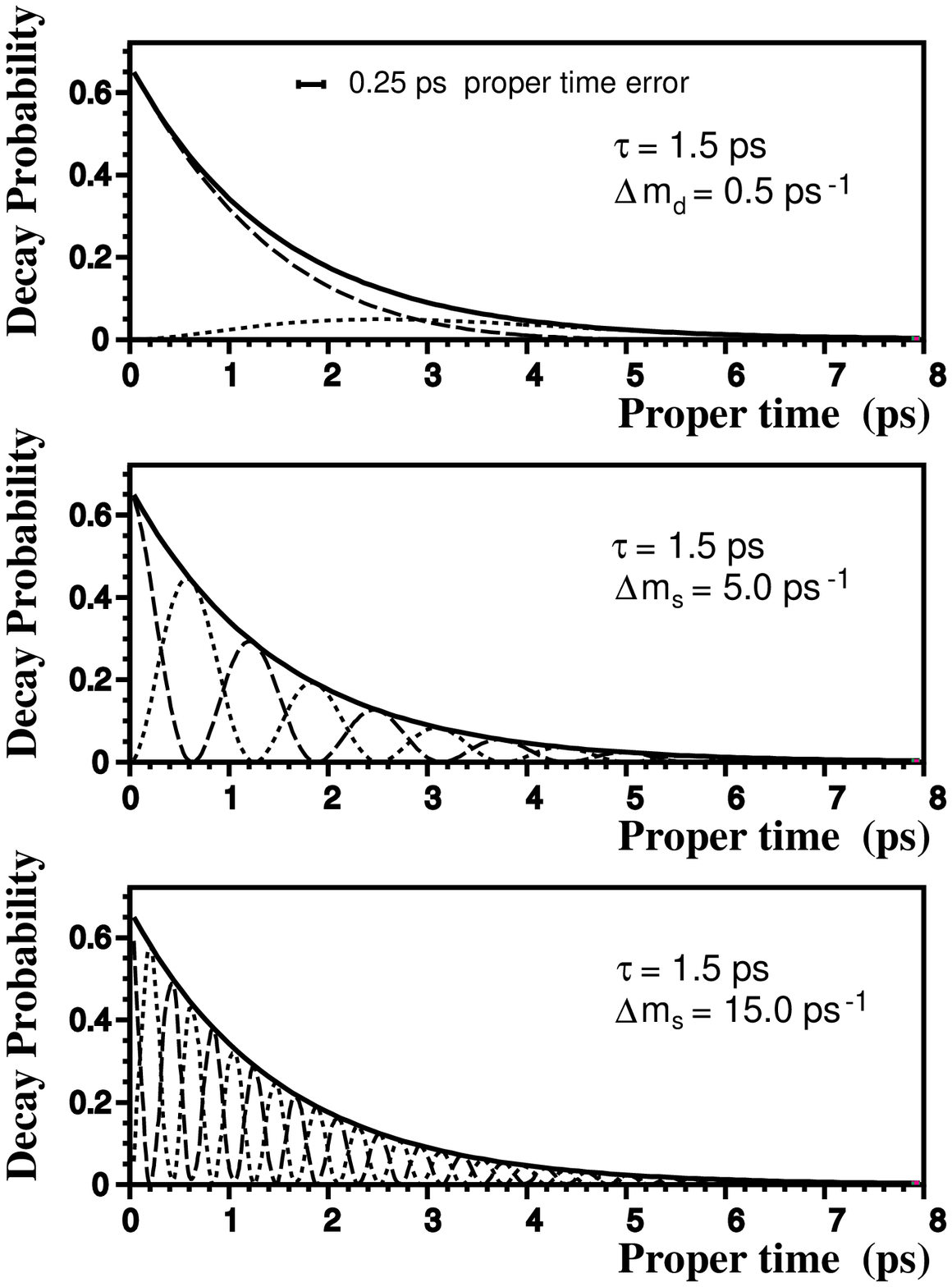,height=4in}
\end{center}
\vspace*{13pt}
\fcaption{${\cal P}_{u}$\/ and ${\cal P}_{m}$\/ for
(a) $\Delta m$ = 0.5 ps$^{-1}$. (b) $\Delta m$ = 5 ps$^{-1}$.
(c) $\Delta m$ = 15 ps$^{-1}$.  The solid line shows the exponential
decay of the $B$\/ meson with a lifetime of 1.5 ps.  The dashed line
shows the ${\cal P}_{u}$\/ distribution, and the dotted line shows the
${\cal P}_{m}$\/ distribution.}
\label{fig:PuPm}
\end{figure}
Fig.~\ref{fig:PuPm} addresses the experimental sensitivity
of such a measurement.  Fig.~\ref{fig:PuPm}(a), (b) and (c) each show
the decay probabilities ${\cal P}_{u}(t)$\/ and ${\cal P}_{m}(t)$\/,
discussed above.  Each figure assumes a $B$\/ lifetime of
1.5 ps, and shows the effect of different values of $\Delta m$ on
${\cal P}_{u}(t)$\/ and ${\cal P}_{m}(t)$\/.
Fig.~\ref{fig:PuPm}(a) shows these
probabilities for \(\Delta m_{d} = 0.5\) ps$^{-1}$\/.
It demonstrates that, since the typical experimental
resolution for the LEP experiments is about 0.25 ps in $B$\/ meson
proper time, it is a relatively easy job to measure
$\Delta m_{d}$\/ due to the large oscillation period.
Fig.~\ref{fig:PuPm}(b) illustrates that if 
\( \Delta m_{\mbox{\scriptsize s}} = 5\) ps$^{-1}$, the oscillation period
is still within a comfortable reach of the experimental
sensitivity.  For
\( \Delta m_{\mbox{\scriptsize s}} = 15\) ps$^{-1}$, Fig.~\ref{fig:PuPm}(c) shows that the
experimental sensitivity for LEP experiments makes it difficult to extract
the frequency of oscillation.

The proper time $t$\/ of the $B$\/ decay is obtained through
\begin{equation}
t = l \left( \frac{m_{B}}{p_{B}} \right)
\end{equation}
where $l$\/ is the decay {\em flight distance} between the $B$\/ production
point and decay point, and $m_{B}$\/ and $p_{B}$\/ are respectively
the mass and momentum of the $B$\/ meson.  The flight distance is
measured with the aid of silicon microvertex detectors which
allow the production and decay vertices to be reconstructed precisely.
The $B$\/ meson decay length is determined by reconstructing a decay
vertex formed from a lepton with high transverse momentum (or $p_{t}$\/)
and a ``charm track''.
The ``charm track'' is formed by combining information from several tracks
which are not consistent with coming from the production point of
the $B$\/ and form a secondary  ``charm vertex''.  (In the case where
a $D^{*}$\/ from $B$\/ decay is fully reconstructed, a variation of
this method is used; see section 3.3).
Because of the presence of tails in the flight distance resolution, it
must be parametrized with several Gaussians.
Typically half of the measurements fall in the ``core'', where the
error is smallest.  This core resolution is $260 \: \mu m$\/ for 
ALEPH\cite{AL:bsljc},
$340 \: \mu m$\/ for DELPHI\cite{DE:bdall},
and $400 \: \mu m$\/ for OPAL\cite{OP:bdsll}.

The $B$ momentum is obtained by reconstructing the momenta of the charged
and neutral decay products
of the $B$\/.  The charged momentum can be reconstructed by
simply summing the momenta of charged tracks consistent with coming from
the decay of the $B$\/.  These usually include a lepton and other charged
tracks from a charm meson decay vertex.  The neutral energy reconstruction
is generally more complicated, involving information from the whole event,
the beam energy, and energy-momentum conservation.  The~$B$\/~momentum
core resolution in ALEPH\cite{AL:bdall}, DELPHI\cite{DE:bdall}
and OPAL\cite{OP:bdsll} is about 8--10\%.

The charge of the $b$\/ quark when it is created (the production flavor)
is typically found in one of
two ways.  Some analyses require a lepton in the hemisphere opposite to
the lepton used to determine the decay flavor, and use its sign to
determine the production flavor.  Others use {\em Jet charge} techniques,
which weight momentum information from charged tracks in the event
to determine the production flavor.

\pagebreak
The charge of the $b$\/ quark when it decays (the decay flavor)
can be measured in a variety of
ways.  In some analyses, the sign of a high $p_{t}$\/ lepton is used to identify
the decay flavor.  In other
analyses, a $D^{*\pm}$\/ is reconstructed, and the
sign of the $D^{*\pm}$\/ is used to determine the $B_{d}^{0}$ decay flavor.

Details of the specific methods used are discussed in the next section.

\section{Measurement of $\Delta m_{d}$\/.}
$B_{d}$\/ oscillation was first observed by ALEPH\cite{AL:Bsone}
at LEP in 1993.
Since then, a vast number of measurements have emerged using a variety
of methods.  The popular methods are described here, namely the
``Lepton--Jet charge'' method, the ``Lepton--Lepton'' method and
methods using a $D^{*}$\/.  The names of these methods are chosen
such that the word before the hyphen refers to the way the
decay flavor is determined, while that after the hyphen is for the
corresponding production flavor.

\subsection{Lepton--Jet charge Method.}
\begin{figure}
\begin{center}
\epsfig{file=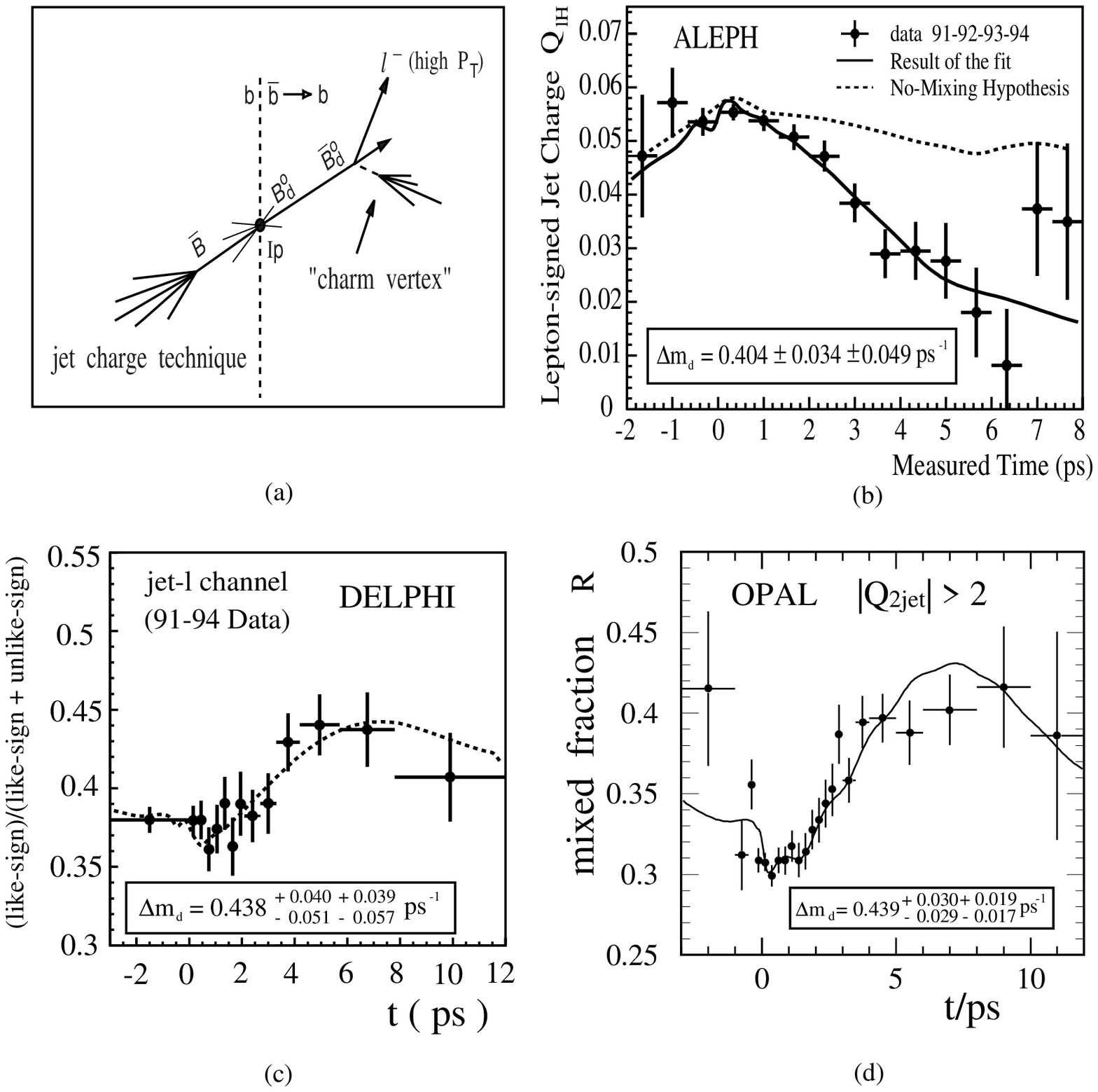,width=6in}
\end{center}
\vspace*{13pt}
\fcaption{Measurement of $\Delta m_{d}$\/ with the Lepton--Jet charge
method. (a) Schematic. (b) ALEPH measurement. (c) DELPHI measurement.
(d) OPAL measurement.}
\label{fig:BdLJC}
\end{figure}

In the Lepton--Jet charge method, events with semileptonic decay
\( b \rightarrow X \ell^{-} \bar{\nu} \)\/
or
\( \bar{b} \rightarrow X \ell^{+} \nu\)\/, %
(\( \ell^{\pm} = e^{\pm} {\it or } \mu^{\pm} \)\/)
are selected with a high $p_{t}$\/ lepton; the charge of
the lepton from these decays identifies the
decay flavor of the $b$\/ quark.  Leptons from
other sources, particularly cascade decays
\( b \rightarrow c \rightarrow \ell^{+} \)\/
dilute the sample of
\( b \rightarrow \ell^{-}\),
but the high $p_{t}$\/ lepton is nevertheless a good measure of
the $b$\/ decay flavor.
The production flavor is then tagged by a Jet charge technique,
discussed below.  The $B$\/ meson decay vertex is determined by the
intersection of the high $p_{t}$\/ lepton and the ``charm track''
on one side of the event, as described in section~2. 
The schematic of this method is shown in Fig.~\ref{fig:BdLJC}(a).

There are several different algorithms used to determine the
$b$\/ production flavor.  The choice of
charged tracks used in these analyses varies.  ALEPH\cite{AL:bdall} and
DELPHI\cite{DE:bdall} use only the tracks in
the hemisphere opposite to the lepton,
while OPAL\cite{OP:bdsljc} uses charged tracks from both the lepton jet,
and the highest energy jet which does not contain the lepton, also called
the {\em opposite side} jet, in calculating
its jet charge.  For convenience, this paper will
refer to both jet charge and hemisphere charge measurements as
jet charge measurements.  

The weighting scheme for the ALEPH and DELPHI results takes the form
\begin{equation}
Q_{H} =  \frac{\sum_{i=1}^{n_{H}} w_{i} q_{i}}
              {\sum_{i=1}^{n_{H}} w_{i}}
\label{eq:Q_H}
\end{equation}
where $n_{H}$\/ is the number of tracks in the opposite hemisphere, $q_{i}$\/
is the charge of the track,
and $w_{i}$\/ is the weight used, taking the form
\( \left| \vec{p}_{i} \cdot \vec{e} \right|^{\kappa}\)\/,
where $\vec{e}$\/ is the direction of the thrust axis for ALEPH,
and the direction of the sphericity
axis for DELPHI. ALEPH calculates this weight using
$\kappa$\/~=~0.5, while DELPHI
uses $\kappa$\/~=~0.6.  OPAL uses a different jet charge,
defined by
\begin{equation}
Q_{2J} = \sum_{i} q_{i}  - 10 \sum_{j} q_{j}\left( \frac{p_{j\parallel}}
                                                   {E_{\rm beam}}
                                            \right)
\label{eq:OPALJC}
\end{equation}
where the first sum is over the tracks in the jet containing the
high $p_{t}$\/ lepton, and the second sum is over the
opposite side jet.  In Eq.~(\ref{eq:OPALJC}), $E_{\rm beam}$\/~is~the beam
energy, and $p_{i\parallel}$\/ is the charged track's momentum
parallel to its jet axis.

Because only one high $p_{t}$\/ lepton is required, the 
``Lepton--Jet charge'' method retains a relatively large sample of
events and hence possesses a strong statistical power.  The
{\em tag rate}\/, or the fraction of events correctly identified
as mixed or unmixed, for jet charge analyses is approximately
70\% for both mixed and unmixed events.  The ALEPH result studies
the time dependence of the lepton-signed jet charge (jet charge multiplied
by sign of lepton) distribution,
without explicitly identifying events as mixed or unmixed.  The results
of $\Delta m_{d}$\/ from ALEPH\cite{AL:bdall}, DELPHI\cite{DE:bdall}
and OPAL\cite{OP:bdsljc} using this method are
shown in Figs.~\ref{fig:BdLJC}(b), \ref{fig:BdLJC}(c) and
\ref{fig:BdLJC}(d).  Throughout this report, where errors are given,
the first is statistical and the second is systematic.

\subsection{Lepton--Lepton Method}
\begin{figure}
\begin{center}
\epsfig{file=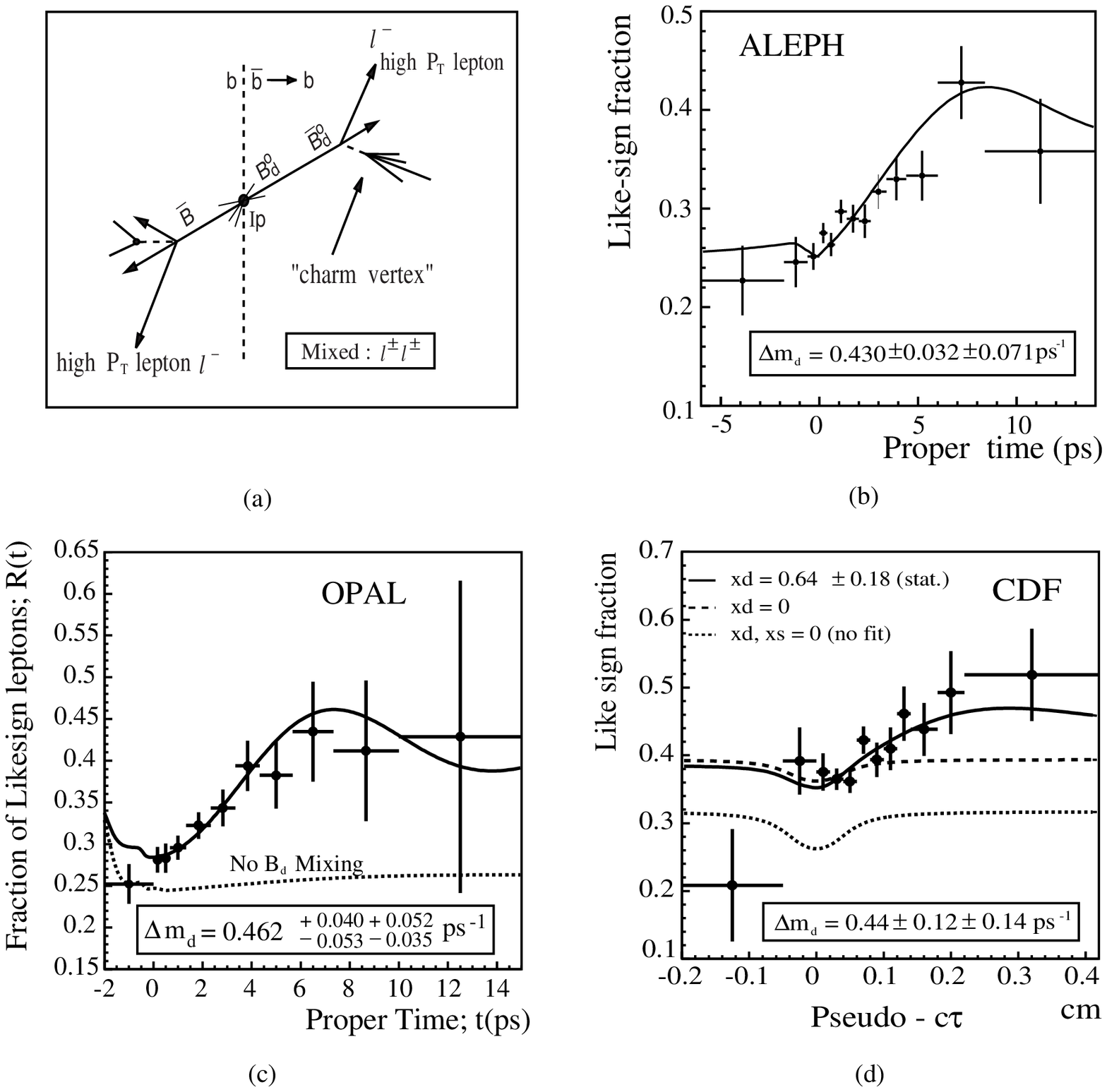,width=6in}
\end{center}
\vspace*{13pt}
\fcaption{Measurement of $\Delta m_{d}$\/ with the Lepton--Lepton
method. (a) Schematic. (b) ALEPH measurement. (c) OPAL measurement.
(d) CDF measurement.}
\label{fig:BdLL}
\end{figure}

Dilepton measurements are in many ways similar to jet charge measurements.
In the Lepton--Lepton method, events with semileptonic decay
\( b \rightarrow X \ell^{-}\bar{\nu} \)\/  or
\( \bar{b} \rightarrow X \ell^{+}\nu \)\/, 
($\ell^{\pm}$~=~$e^{\pm}$~or~$\mu^{\pm}$\/), on both sides of the event are
selected.  The $B$\/ meson decay vertex is determined by the intersection
of the high $p_{t}$\/ lepton and the ``charm track'' on one side of
the event, as described in section~2.  The $B$\/ meson
decay flavor is tagged by the sign of the high $p_{t}$\/ lepton on
the flight distance side of the event, just as in the
Lepton--Jet charge method.
The flavor at production time is tagged by a lepton in the
opposite hemisphere.  Contributions from mixing of the opposite side
$B$\/ hadron are independent of the proper time in the flight distance
hemisphere, and their effect is factored into the tag rate calculation.
The entire process can then be repeated with
the roles of the leptons reversed, giving up to two measurements per
event.  The schematic of this method is shown in Fig.~\ref{fig:BdLL}(a).

Because of the requirement that a lepton be found in each hemisphere,
Lepton--Lepton measurements have a smaller event sample than
Lepton--Jet charge analyses.  Compensating for their smaller event
sample, Lepton--Lepton analyses have superior
tag rates, correctly identifying events as mixed or unmixed
80\% of the time.  This gives
them sensitivity comparable to the Lepton--Jet charge method.
The results from ALEPH\cite{AL:bdall}, 
OPAL\cite{OP:bdsll}, and CDF\cite{CDF:bdll} using this method
are presented in Fig.~\ref{fig:BdLL}(b),
\ref{fig:BdLL}(c), and \ref{fig:BdLL}(d).

Using the Lepton--Lepton method, the DELPHI experiment\cite{DE:bdall} finds
$\Delta m_{d} = 0.42 \pm 0.08^{+0.08}_{-0.07}$~ps$^{-1}$.
DELPHI extends this method by including the use of a charged kaon
to identify the decay flavor of the $B$\/ meson, making use of DELPHI's
unique feature, the RICH counters.  Thus, in the flight distance hemisphere,
the measurement uses a lepton or a charged kaon coming from the secondary
vertex, relying on the dominant decay~chain
\( b \rightarrow c \rightarrow s \)\/ to identify the $B$\/ flavor.
Such a kaon can be used in either the flight distance hemisphere to
determine the decay flavor, or the opposite hemisphere, to determine
the production flavor.  This analysis also incorporates the jet charge
or lepton in the opposite hemisphere to determine the
production flavor.  The DELPHI measurement using this
Lepton--Kaon--Jet charge method\cite{DE:bdall} is
$\Delta m_{d} = (0.563^{+0.050}_{-0.046}\pm0.058)$~ps$^{-1}$.
Because of the strong statistical correlation
expected between this result and the
other inclusive DELPHI results, it has not been included in the
final $\Delta m_{d}$\/ average given in Section~3.4.

\subsection{Methods using a $D^{*}$\/.}
\begin{figure}
\begin{center}
\epsfig{file=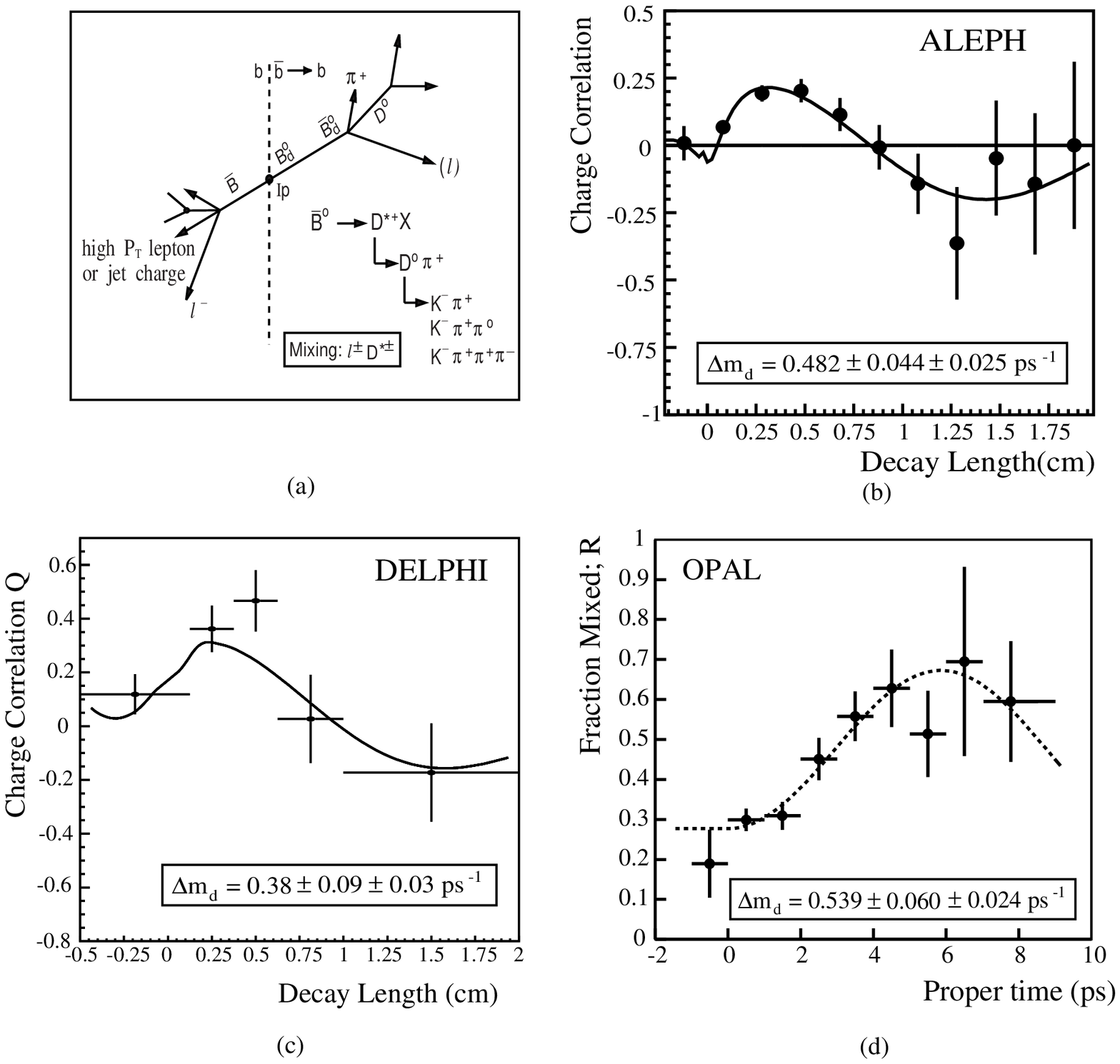,width=6in}
\end{center}
\vspace*{13pt}
\fcaption{Measurement of $\Delta m_{d}$\/ with the $D^{*}$ methods.
(a) Schematic. (b) ALEPH $D^{*}$\/--Lepton or Jet charge measurement.
(c) DELPHI $D^{*}$~Lepton--Jet charge measurement.
(d) OPAL $D^{*}$~Lepton--Jet charge measurement.

The combined $D^{*}$\/ based result from DELPHI is $\Delta m_{d} = 0.421\pm 0.064\pm
0.042$\/ ps$^{-1}$\/.}
\label{fig:BdDL}
\end{figure}
It is also possible to carry out time-dependent measurements of
$\Delta m_{d}$\/
by reconstructing a $D^{* \pm}$\/ from the decay of a $B$\/ meson.
By selecting a charged $D^{*}$\/ sample, it is possible to obtain a
very pure sample of $B_{d}$\/ mesons.  Though some $B^{+}$\/ mesons
contribute to the $D^{* -}$\/ sample, the contamination is small,
and the effect of
these $B^{+}$\/ decays can be included in the fit for $\Delta m_{d}$\/.
Because a $D^{*}$\/
candidate must be reconstructed, these analyses typically have much
smaller event samples than either the Lepton--Jet charge or
the Lepton--Lepton measurements.  Two methods are described here, the
``$D^{*}$\/--Lepton or Jet charge'' method, and the
``$D^{*}$\/ Lepton--Jet charge'' method.  The schematic for these
methods is shown in Fig.~\ref{fig:BdDL}(a).

In the $D^{*}$\/--Lepton or Jet charge analyses, a $D^{0}$\/ sample is
reconstructed using the decays to $K\pi$\/, $K\pi\pi^{0}$, and
$K\pi\pi\pi$\/.
and then the $D^{0}$\/ is combined with a pion to
produce a charged $D^{*}$\/.  The decay flavor of the $B_{d}$\/ is
identified by the sign of this $D^{*}$\/.  The production flavor
can be identified
with either a jet charge technique or with a lepton in the hemisphere
opposite the $D^{*}$\/, as discussed in the previous sections.
As the pion from the $D^{*}$\/ decay has low momentum, and travels
along the flight direction, it cannot be used to determine the
$B$\/ decay point.  The apparent $D^{0}$\/ decay vertex is used
to infer the $B$\/ flight distance.
The result from the ALEPH experiment\cite{AL:bdall} using the
$D^{*}$\/--Lepton or Jet charge method is shown in
Fig.~\ref{fig:BdDL}(b).  The DELPHI experiment\cite{DE:bdall} obtains 
$\Delta m_{d}$~=~0.470~$\pm$~0.086~$\pm$~0.061~ps$^{-1}$ with this
method, while OPAL\cite{OP:bddsl} finds
$\Delta m_{d}$~=~0.57~$\pm$\/~0.11$\pm$\/0.02 ps$^{-1}$\/.

In the $D^{*}$\/ Lepton--Jet charge analyses, a $D^{*}$\/ sample is
produced as described above, and a lepton in the same hemisphere
is used to form a $B_{d}$ decay vertex, for measuring the flight
distance.  The decay flavor is determined by the sign of the $D^{*}$\/,
and the production flavor is determined by a jet charge technique,
as discussed in previous sections.  The result from the DELPHI
$D^{*}$\/~Lepton--Jet charge analysis\cite{DE:bdall} is shown in
Fig.~\ref{fig:BdDL}(c), while the OPAL result\cite{OP:bddsljc} is shown in
Fig.~\ref{fig:BdDL}(d).
\pagebreak
DELPHI has performed an average of their $D^{*}$ based analyses\cite{DE:bdall},
giving $\Delta m_{d} = (0.421\pm 0.064\pm 0.042)$\/ ps$^{-1}$,
which has been included in the LEP and world averages computed in
Section~3.4.

\subsection{Average of $\Delta m_{d}$\/ Results.}
\begin{figure}
\begin{center}
\epsfig{file=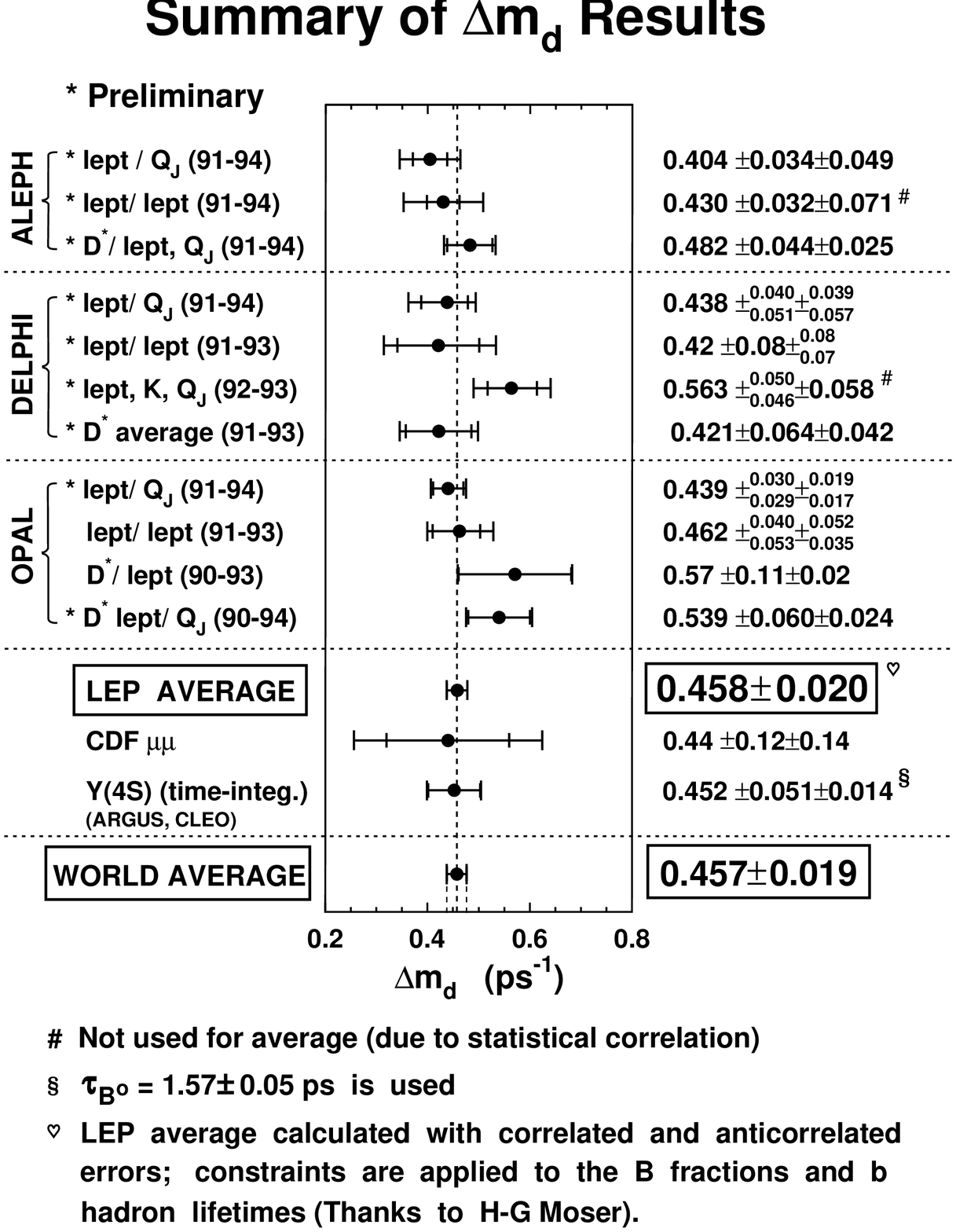,width=6in}
\end{center}
\vspace*{13pt}
\fcaption{Summary of results for $\Delta m_{d}$\/.}
\label{fig:dmd}
\end{figure}
The process of performing an average on the measurements of $\Delta m_{d}$\/
is complicated by the presence of correlations between the various
measurements. These
correlations can be {\em statistical}\/, coming from overlapping
data samples, or {\em systematic}\/, coming from common assumptions
used in making the measurement.

Where measurements are statistically correlated, the degree of
correlation is generally unknown.  To minimize correlations, when there
have been multiple
measurements of similar types performed on the same data sample, the
average includes only those results with the smallest errors.
Where measurements are
systematically correlated, it is possible to judge the degree of
correlation between different results by looking at common correlated
systematic errors.

The main correlated systematic errors come from the lifetimes and
fractions of individual $B$\/ hadron species.  Thus, the averaging
process
considers correlations related to the lifetimes and fractions only.
There are other errors which are correlated, in theory, such as the
parametrization of the $b$\/ fragmentation function, and the
decay length resolution in individual LEP experiments, but these
errors are generally smaller, and their correlations can safely be
neglected.

Taking these correlations into account correctly is complicated
by the fact that each measurement parametrizes these systematic
effects in a different way, and use different central values and
errors on their parameters.  The average considers the individual
parametrizations of these different experiments, and performs a
constrained fit\cite{AL:Moser} for $\Delta m_{d}$\/, the lifetimes and the
$B$\/ hadron fractions.

Using this averaging technique, and excluding the ALEPH lepton--lepton
measurement and the DELPHI Lepton--Kaon--Jet charge measurement due
to statistical overlap, the LEP average is found to be
$\Delta m_{d}$~=~0.458~$\pm$~0.020~ps$^{-1}$.  Including results
from CDF and time integrated measurements from $\Upsilon$\/(4s) in the average yields
$\Delta m_{d}$ = 0.457 $\pm$ 0.019 ps$^{-1}$.  A summary of the
results is presented in Fig.~\ref{fig:dmd}.

\section{Measurement of $\Delta m_{\mbox{\scriptsize s}}$\/.}
\begin{figure}
\begin{center}
\epsfig{file=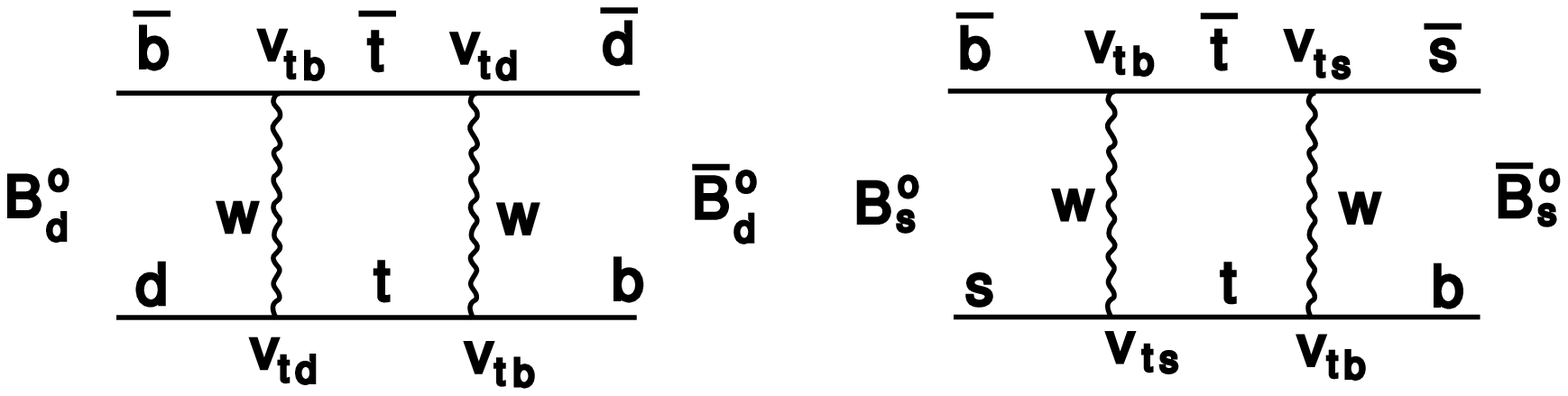,width=6in}
\end{center}
\fcaption{Important quark diagrams (including those with $W$\/ and top quark lines exchanged) for the calculation of
$\Delta m_{d}$\/ and $\Delta m_{\mbox{\scriptsize s}}$\/.}
\label{fig:Quark}
\end{figure}
Both $\Delta m_{d}$\/ and $\Delta m_{\mbox{\scriptsize s}}$\/ are mass differences between
particles and hence they are of direct physical importance.  A further
motivation
for measuring these quantities comes from the fact that these mass
differences are due to high-order weak interactions.  The important
diagrams for these interactions are shown in Fig.~\ref{fig:Quark}
together with similar diagrams with the $W$\/
and top quark lines exchanged.  Computation from these diagrams gives
\begin{equation}
\frac{\Delta m_{\mbox{\scriptsize s}}}     {\Delta m_{d}} \simeq
\frac{       m_{B_{\mbox{\scriptsize s}}}} {       m_{B_{d}}}
\left|
\frac{V_{ts}}{V_{td}}
\right|^{2}
\xi_{\mbox{\scriptsize s}}^{2}
\frac{\hat{\eta}_{B_{\mbox{\scriptsize s}}}}
     {\hat{\eta}_{B_{d}}}
\label{eq:qcalc}
\end{equation}
where $\hat{\eta}_{B_{\mbox{\scriptsize s}}}$\/ and $\hat{\eta}_{B_{d}}$\/ are the QCD
correction factors for the $B_{\mbox{\scriptsize s}}$\/ and $B_{d}$, expected to be
similar, and $\xi_{\mbox{\scriptsize s}}$\/ is the ratio of hadronic matrix elements for
the $B_{\mbox{\scriptsize s}}$\/ and $B_{d}$\/.  Estimates from lattice QCD\cite{th:lattice}
and QCD sum rules\cite{th:sum} are consistent with a value\cite{th:comb}
of $\xi_{\mbox{\scriptsize s}}$~=~1.16~$\pm$~0.10.  Measurements of
$\Delta m_{d}$\/
and $\Delta m_{\mbox{\scriptsize s}}$\/ can therefore be used to determine
\( \left| V_{ts}/V_{td} \right| \)\/.
This ratio of the CKM matrix elements is of special interest because it
appears in one of the most useful unitarity triangles given by
\begin{equation}
\frac{V_{td}}{V_{ts}}    + 
V_{us}^{*}               + 
\frac{V_{ub}^{*}}{V_{ts}}  =  0.
\end{equation}

For a number of reasons, the measurement of $\Delta m_{\mbox{\scriptsize s}}$\/ is much
more difficult than that for $\Delta m_{d}$\/.  The theoretical
expectation that $\Delta m_{\mbox{\scriptsize s}}$\/ is significantly larger than
$\Delta m_{d}$\/ has been confirmed by
experiments\cite{OP:bdsll,AL:Forty}.
As illustrated
in Fig.~\ref{fig:PuPm}, such larger values of $\Delta m_{\mbox{\scriptsize s}}$\/ lead to rapid
oscillation, which complicates the measurement.  A second difficulty
comes from the fact that $f_{B_{\mbox{\scriptsize s}}}$\/, the fraction of $B_{\mbox{\scriptsize s}}$\/ 
produced in $b$\/ decays, is substantially smaller than
$f_{B_{d}}$\/ and is not well measured.  

Three methods for determination of $\Delta m_{\mbox{\scriptsize s}}$\/ will be
described here.  They are
the ``Lepton--Jet charge'' method, the ``Lepton--Lepton'' method
and the ``Lepton--Kaon Correlation'' method.

\subsection{Lepton--Jet charge Method}
This method has been used by the ALEPH\cite{AL:bsljc},
DELPHI\cite{DE:bdall},
and OPAL\cite{OP:bdsljc} collaborations
at LEP, and its schematic is that of Fig.~\ref{fig:BdLJC}(a)
with $B_{d}$\/ and
$\bar{B}_{d}$\/ replaced by $B_{\mbox{\scriptsize s}}$\/ and $\bar{B}_{\mbox{\scriptsize s}}$\/.  Again, the
$B$\/ decay vertex is formed by the secondary vertex including a
high $p_{t}$\/ lepton.  The decay flavor is tagged by the high $p_{t}$\/
lepton while the production flavor is tagged by the jet charge technique.

Since the lower bound from ALEPH\cite{AL:bsljc} using this method is the best
one for $\Delta m_{\mbox{\scriptsize s}}$\/, it will be described in detail here.
The main difference between the ALEPH Lepton--Jet charge method for
$\Delta m_{\mbox{\scriptsize s}}$\/ and that for $\Delta m_{d}$\/ described in Sec.~3.1 is
that a different jet charge algorithm is used.  Instead of using the
$Q_{H}$\/ in
Eq.~(\ref{eq:Q_H}) where only the charged tracks in the opposite hemisphere are
used, the new jet charge algorithm makes use of information from both the
opposite jet and the flight distance jet.  The weight applied to each
track in computing the jet charge value for the event is the track's
rapidity with respect to the jet axis.  More precisely, define
\begin{figure}[tb]
\begin{center}
\epsfig{file=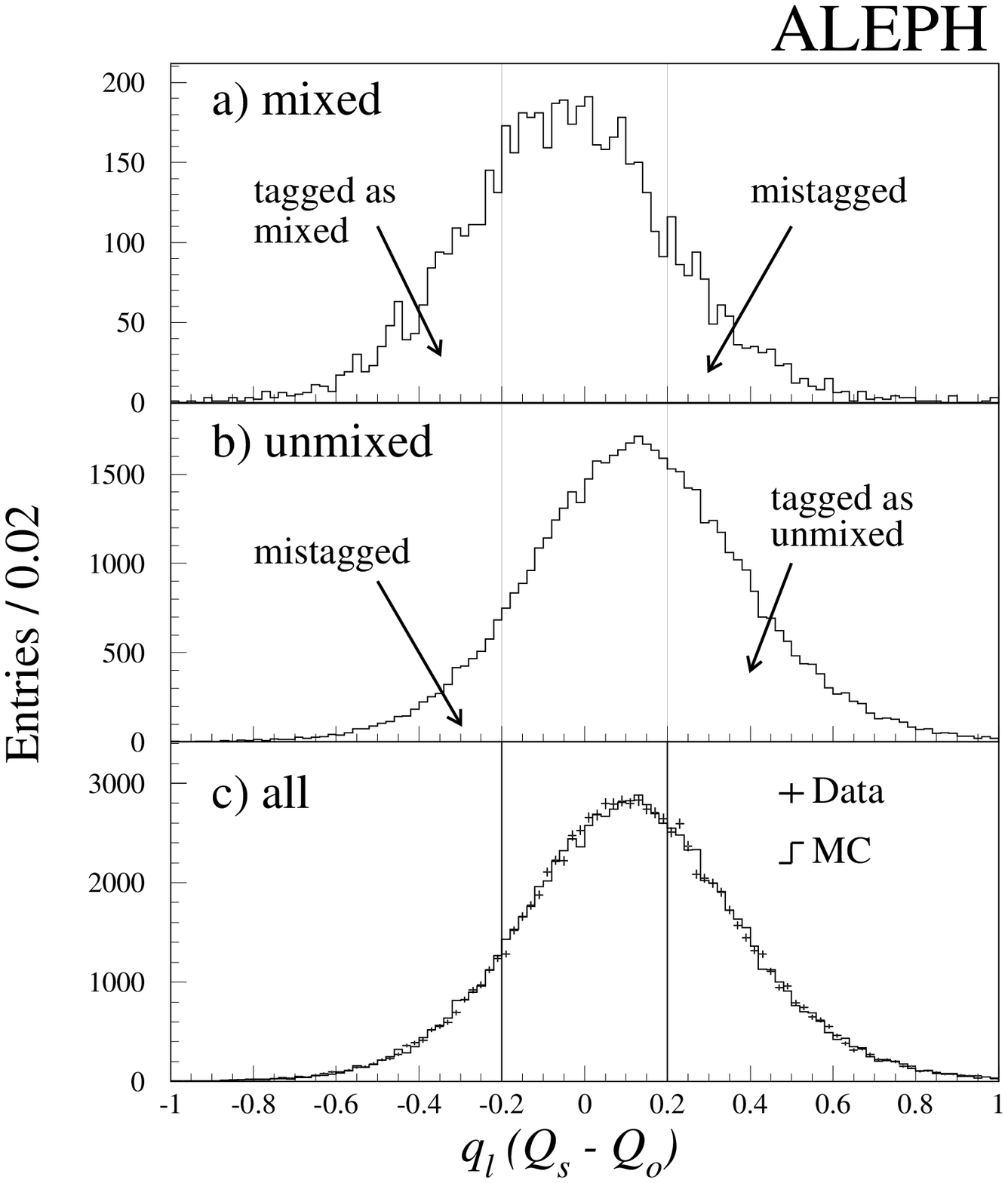,height=4in}
\end{center}
\fcaption{Lepton signed same side minus opposite side jet charge
distributions for a) mixed $B^{0}$\/ mesons, and b) unmixed $b$\/
hadrons in the full Monte Carlo simulation.  Plotted in c) is the
sum of all Monte Carlo events, normalized to the number of events
in the data.  The data points are superimposed with error bars.  The
vertical lines indicate the selection requirement
$|Q_{S} - Q_{O}| > 0.2$.}
\label{fig:qlQ2}
\end{figure}
\begin{equation}
Q_{S,O} = \frac{ \sum_{i} y_{i} q_{i}}
               { \sum_{i} y_{i}}
\end{equation}
where $S$\/ and $O$\/ indicate the sum is over tracks in the same side jet
(the jet with the high $p_{t}$\/ lepton) and opposite side jet, respectively.  The
rapidity, $y_{i}$\/, is given by
\begin{equation}
y_{i} = \frac{1}{2} \ln{\frac{E_{i} + P_{i \parallel}}
                             {E_{i} - P_{i \parallel}}}.
\end{equation}
The jet charge variable used to identify the production flavor is then
\begin{equation}
Q = Q_{S} - Q_{O}.
\label{eq:Qfinal}
\end{equation}

For the purpose of tagging mixed or unmixed events, the
charge $q_{\ell}$\/ of the high $p_{t}$\/ lepton and the sign of the
above $Q$\/ are used.  Events are identified as follows:
\begin{displaymath}
\begin{array}{ccccr}
+ \: - & , & - \: + & \longleftrightarrow & {\it mixed \: events} \nonumber \\
+ \: + & , & - \: - & \longleftrightarrow & {\it unmixed \: events} \nonumber
\end{array}
\end{displaymath}
As shown in Fig.~\ref{fig:qlQ2}, a cut requiring 
\( \left| q_{\ell} \cdot Q \right| > 0.2\)\/ is imposed.
With this cut, the tag rate for unmixed events, $A_{u}$\/, is about
80\% and for mixed events, $A_{m}$\/, is about 60\%.  Since there are
eight times more unmixed events than mixed events, it is essential to
have a high tag rate for unmixed events.  This is a great advantage
of using the $Q$\/ defined in Eq.~(\ref{eq:Qfinal}).  To make optimal
use of the experimental data, the method of maximum likelihood is used
to extract the value of $\Delta m_{\mbox{\scriptsize s}}$\/.

\begin{figure}[tb]
\begin{center}
\epsfig{file=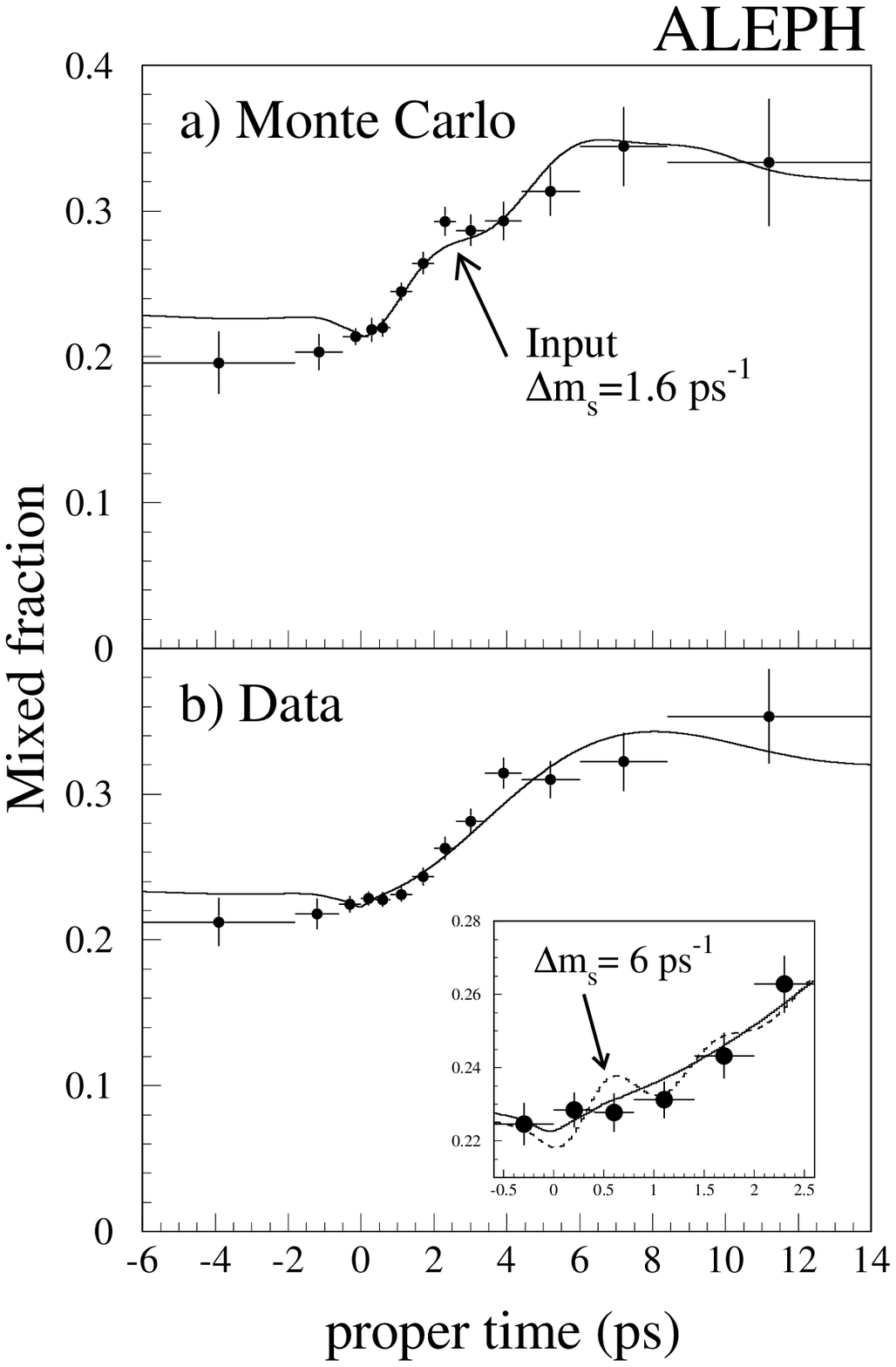,height=4in}
\end{center}
\fcaption{The tagged mixed fraction of events as a function of measured
proper time, for a) Monte Carlo with $\Delta m_{\mbox{\scriptsize s}}$\/~=~1.6~ps$^{-1}$\/,
and b) data events.  The superimposed curve for the Monte Carlo in
a) is the expected distribution for $\Delta m_{\mbox{\scriptsize s}}$~=~1.6~ps$^{-1}$.
The solid curve for the data in b) assumes $\Delta m_{\mbox{\scriptsize s}}$~=~30~ps$^{-1}$,
while the dashed curve in the insert is the expected distribution for
$\Delta m_{\mbox{\scriptsize s}}$~=~6~ps$^{-1}$.  The small proper time region of the
plot is expanded to emphasize the part most sensitive to $B_{\mbox{\scriptsize s}}$\/
oscillations.}
\label{fig:tagmix}
\end{figure}
\begin{figure}[p]
\begin{center}
\epsfig{file=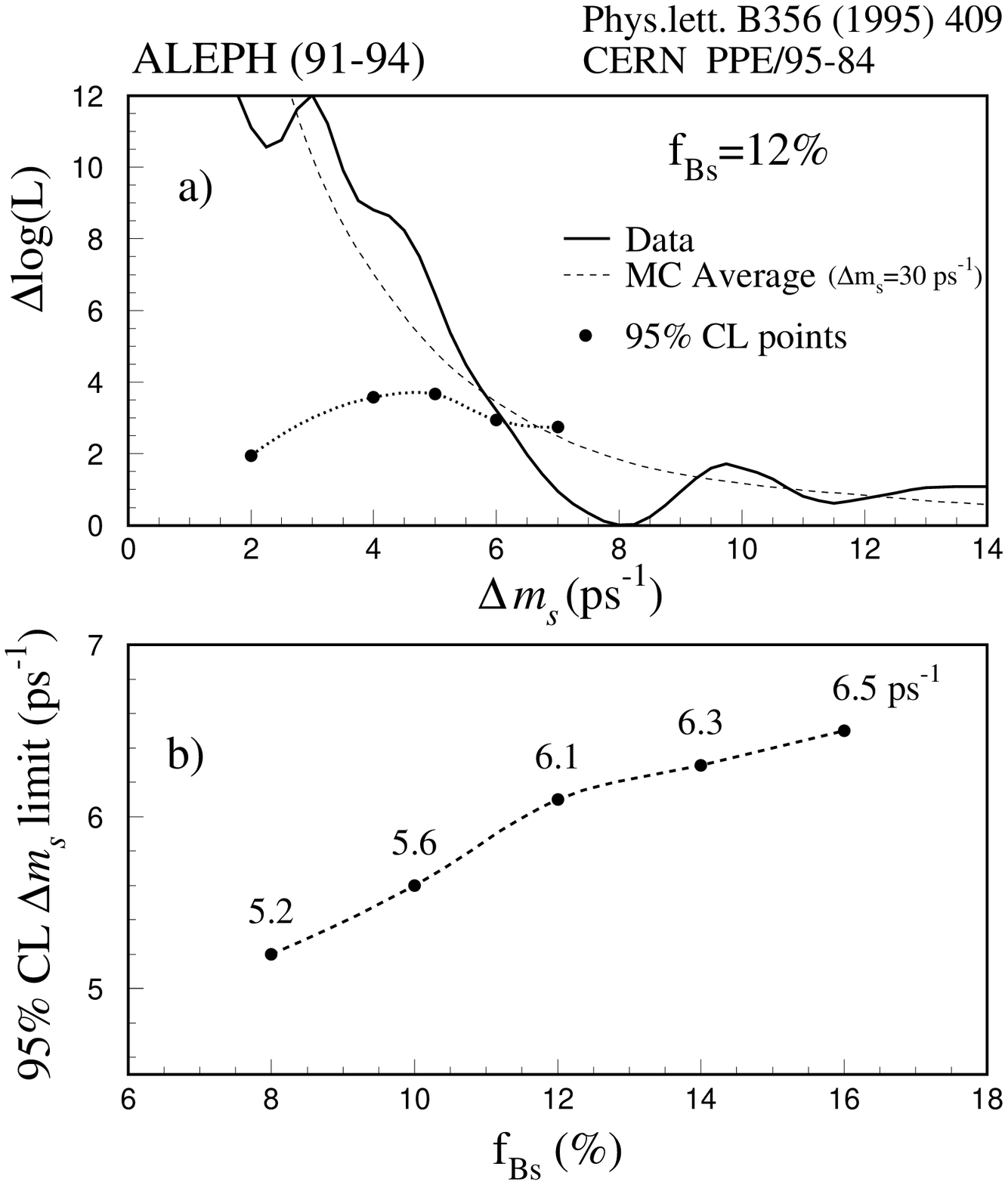,height=5in}
\end{center}
\fcaption{Superimposed on the data likelihood curve are the average of
the fast Monte Carlo $\Delta \log{L}$\/ values, and the 95\% confidence
limit points.  $\Delta \log{L}$\/ is defined as the ($-\log{L}$) value
at the $\Delta m_{\mbox{\scriptsize s}}$\/ value, minus the minimum ($-\log{L}$) value.  A
limit curve is drawn through the 95\% confidence limit points.  The
data curve crosses the limit curve at $\Delta m_{\mbox{\scriptsize s}}$~=~6.1~ps$^{-1}$
for $B_{\mbox{\scriptsize s}}$\/ fraction $f_{B_{\mbox{\scriptsize s}}}$~=~12\%.  Also shown is the average
of the $\Delta\log{L}$\/ curves for 200 fast Monte Carlo samples in
which $B_{\mbox{\scriptsize s}}$\/ mixing is near-maximal ($\Delta m_{\mbox{\scriptsize s}}$~=~30~ps$^{-1}$\/).
b) the results of 95\% confidence level limit in $\Delta m_{\mbox{\scriptsize s}}$\/
as a function of the $B_{\mbox{\scriptsize s}}$\/ fraction $f_{B_{\mbox{\scriptsize s}}}$.  The limits for
$\Delta m_{\mbox{\scriptsize s}}$\/ are 5.2, 5.6, 6.1, 6.3, and 6.5 for $f_{B_{\mbox{\scriptsize s}}}$~=~8\%,
10\%, 12\%, 14\% and 16\%, respectively.}
\label{fig:BsLimit}
\end{figure}
Fig.~\ref{fig:tagmix} shows the tagged mixed fraction for Monte Carlo and data.  The value of
$\Delta m_{d}$\/ is determined as a check of the tag rates
used in this analysis, and of the peformance
of the likelihood fit.  Assuming a $B$\/ meson lifetime,
$\tau_{B}$\/, of 1.5 ps, $B_{d}$\/
fraction, $f_{B_{d}}$\/, of 0.4,
$B_{\mbox{\scriptsize s}}$\/ fraction,
$f_{B_{\mbox{\scriptsize s}}}$\/, of 0.12, $\Delta m_{\mbox{\scriptsize s}}$\/ of 30 ps$^{-1}$\/,
and $A_{m}$\/ of 0.6,
a two-dimensional fit with the ALEPH data is performed for $\Delta m_{d}$\/
and $A_{u}$\/.  This gives
\( A_{u}        = 0.792 \pm 0.003 \)\/ and 
\( \Delta m_{d} = 0.47  \pm 0.04\) ps$^{-1}$, the latter value being in
agreement with the world average of Fig.~\ref{fig:dmd}.  Similar fits
with Monte Carlo give
\( A_{u}        = 0.792 \pm 0.003 \)\/ and
\( \Delta m_{d} = 0.48  \pm 0.05\) ps$^{-1}$
to be compared with the input values of
0.790 and
0.467 ps$^{-1}$\/ respectively.

\begin{figure}[tb]
\graftwos {\peteheight},%
{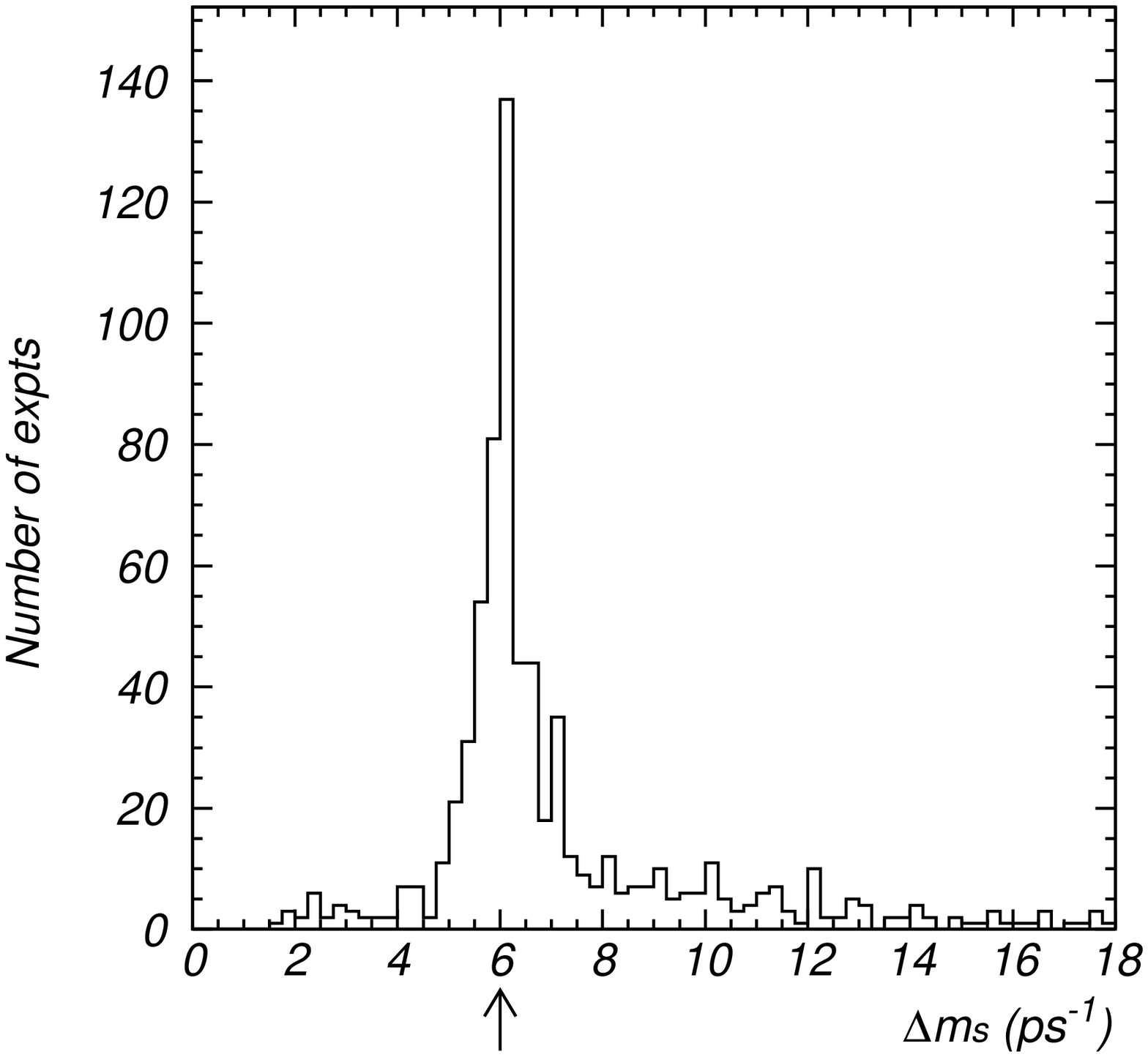},%
{\fcaption
{\label{fig:BsSens}
 The likelihood minima for an input $\Delta m_{\mbox{\scriptsize s}}$\/ = 6 ps$^{-1}$.
}},%
{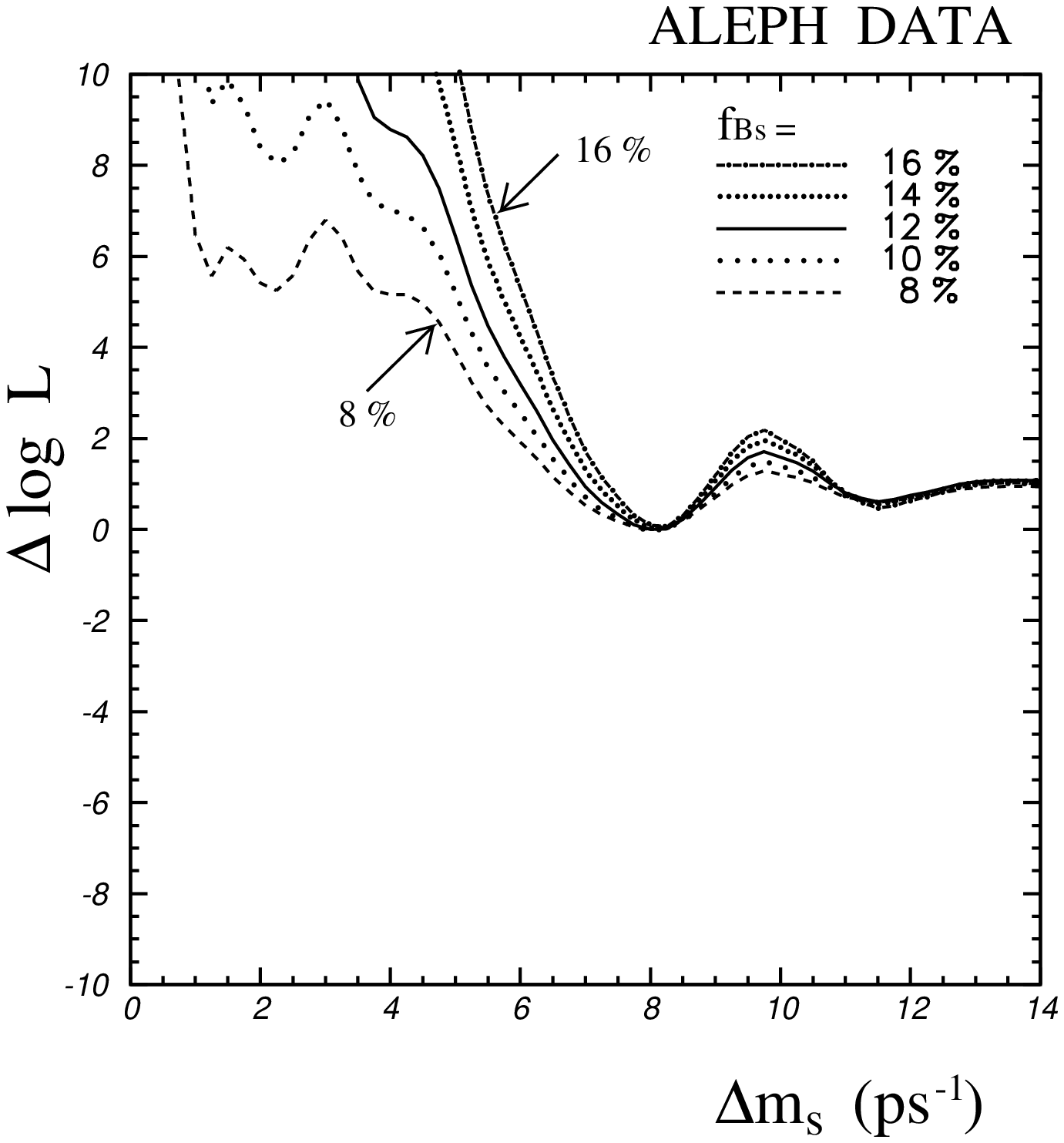},%
{\fcaption
{\label{fig:BsFrac}
 ALEPH data curve for different values of $f_{B_{\mbox{\scriptsize s}}}$.
}}
\end{figure}
The ALEPH result is shown in Fig.~\ref{fig:BsLimit}.  This figure shows the
$\Delta \log{L}$\/ curve for the data as a function of $\Delta m_{\mbox{\scriptsize s}}$\/,
where $\Delta \log{L}$\/ is defined as the negative log likelihood value
($-\log{L}$\/) at a given $\Delta m_{\mbox{\scriptsize s}}$\/ minus the ($-\log{L}$\/) value
calculated at the $\Delta m_{\mbox{\scriptsize s}}$\/ where the ($-\log{L}$\/) is at its
minimum.  It uses the values of $A_{u}$\/ and $\Delta m_{d}$\/ determined
above as inputs to the fit, and assumes a $B_{\mbox{\scriptsize s}}$\/ fraction of
$f_{B_{\mbox{\scriptsize s}}} = 12\%$\/.  The data prefer high values of $\Delta m_{\mbox{\scriptsize s}}$\/,
with a favored value of 8 ps$^{-1}$\/.  The difference in likelihood
for higher values of $\Delta m_{\mbox{\scriptsize s}}$\/ is insufficient to exclude them,
therefore a lower limit is set on $\Delta m_{\mbox{\scriptsize s}}$\/.
Superimposed on the data is a 95\% confidence level lower limit curve
calculated using a `fast' Monte Carlo.

In constructing the limit curve, the likelihood differences,
$\Delta \log{L}$\/, for the fast Monte Carlo
are calculated for 300 samples at various input values of $\Delta m_{\mbox{\scriptsize s}}$\/
(2.0, 4.0, 5.0, 6.0 and 7.0 ps$^{-1}$\/), each with sample size equal
to that of the data.  If the $\Delta m_{\mbox{\scriptsize s}}$\/ value is close to the
point where the limit is set, 600 samples are used.  The 95\% confidence
limit is determined
by locating the point below which lie 95\% of the $\Delta \log{L}$ values,
calculated at the input value of $\Delta m_{\mbox{\scriptsize s}}$\/.  The 95\% confidence
limit curve is then drawn through the points at different input
$\Delta m_{\mbox{\scriptsize s}}$\/, as shown in Fig.~\ref{fig:BsLimit}(a).  The data
$\Delta \log{L}$\/
curve intersects the limit curve at $\Delta m_{\mbox{\scriptsize s}}$\/ = 6.1 ps$^{-1}$\/.
This point is taken as the 95\% confidence level lower limit.
The lower plot of
Fig.~\ref{fig:BsLimit}(b) shows the result of performing this complete
analysis with several different values of $f_{B_{\mbox{\scriptsize s}}}$\/ as discussed
later in this section.

It is important to check that there is indeed sensitivity at
$\Delta m_{\mbox{\scriptsize s}}$\/ = 6 ps$^{-1}$\/.  For this purpose, 800 Monte Carlo
samples were generated at this value of $\Delta m_{\mbox{\scriptsize s}}$\/ with the
statistics of each sample again matching those of the ALEPH data.  For each
of these 800 samples, the value of $\Delta m_{\mbox{\scriptsize s}}$\/ at the minimum of
the ($-\log{L}$\/) curve is determined.  The distribution of these minima are
shown in Fig.~\ref{fig:BsSens}.  The figure clearly shows that in the
majority of cases, the method does find the correct minimum, which
demonstrates that there is indeed sensitivity at
$\Delta m_{\mbox{\scriptsize s}}$\/ = 6 ps$^{-1}$\/.

The data curve of Fig.~\ref{fig:BsLimit}(a) corresponds to an
input $f_{B_{\mbox{\scriptsize s}}}$\/ of 12\%. Fig.~\ref{fig:BsFrac} shows the corresponding
data curves obtained from the ALEPH data for various assumed
$B_{\mbox{\scriptsize s}}$\/ fractions.  The figure demonstrates that the sensitivity
to $\Delta m_{\mbox{\scriptsize s}}$\/ increases as $f_{B_{\mbox{\scriptsize s}}}$\/ increases.
Fig.~\ref{fig:BsLimit}(b) shows the results of 95\% confidence level
lower limit in $\Delta m_{\mbox{\scriptsize s}}$\/ as a function of $f_{B_{\mbox{\scriptsize s}}}$\/.  The
limit varies from $\Delta m_{\mbox{\scriptsize s}} > 5.2$\/ ps$^{-1}$\/ at
$f_{B_{\mbox{\scriptsize s}}}$~=~8\% to $\Delta m_{\mbox{\scriptsize s}} > 6.5$\/ ps$^{-1}$\/ at
$f_{B_{\mbox{\scriptsize s}}}$~=~16\%.

\begin{figure}[t]
\begin{center}
\vspace*{7pt}
\epsfig{file=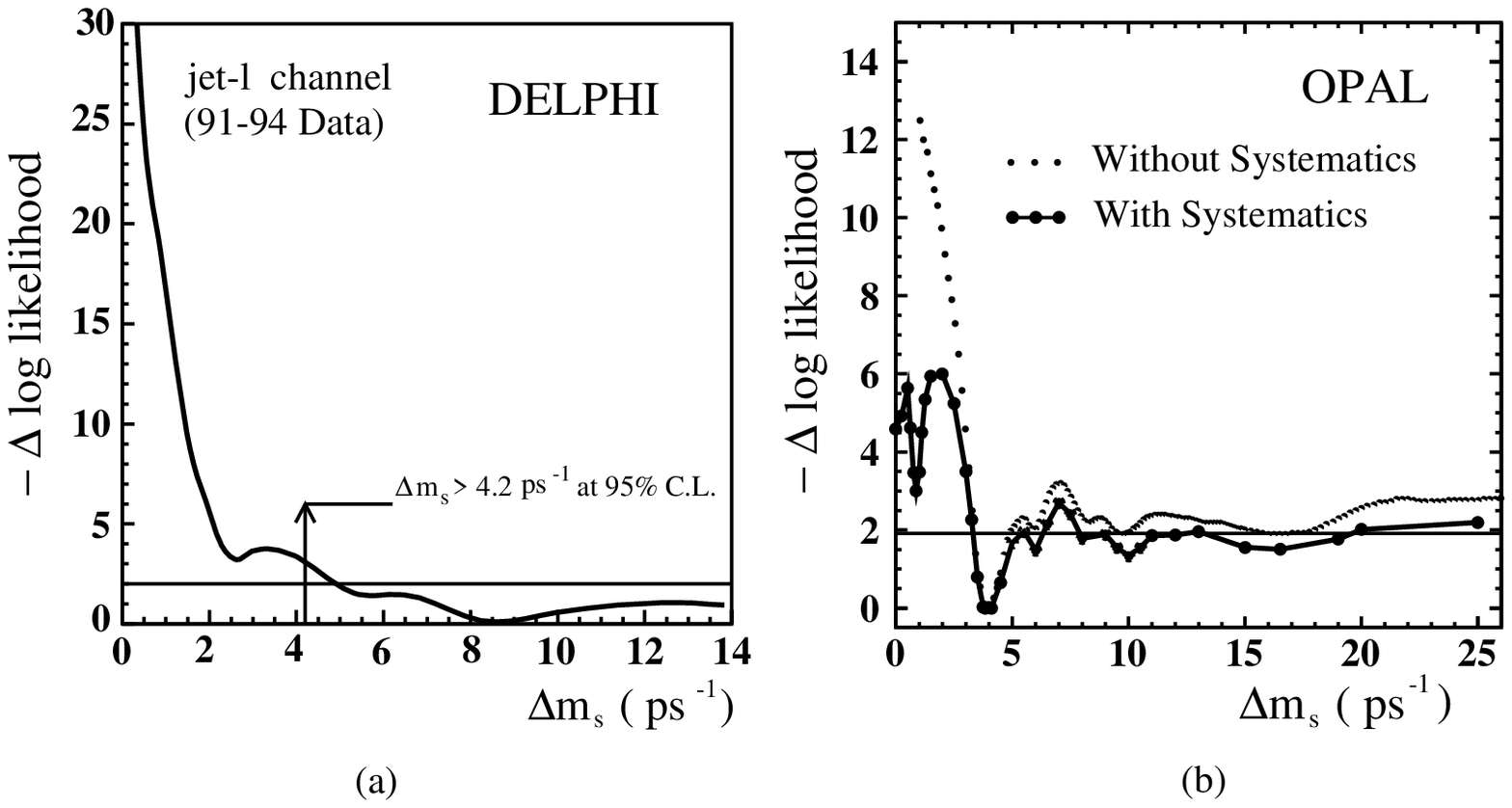,width=6in}
\end{center}
\fcaption{Limit on $\Delta m_{\mbox{\scriptsize s}}$\/ from (a) DELPHI and (b) OPAL
Lepton--Jet charge method.}
\label{fig:BsLJCOther}
\end{figure}
The corresponding preliminary results using the
Lepton--Jet charge method from DELPHI\cite{DE:bdall}
and OPAL\cite{OP:bdsljc} are shown in Fig~\ref{fig:BsLJCOther}.
The result from DELPHI is $\Delta m_{\mbox{\scriptsize s}}>4.2$~ps$^{-1}$ at 95\%~Confidence
Level for $f_{B_{\mbox{\scriptsize s}}}$~=~(10~$\pm$~3)\%, while from OPAL, the result is
$\Delta m_{\mbox{\scriptsize s}}>3.3$~ps$^{-1}$\/ at 95\%~Confidence Level for 
$f_{B_{\mbox{\scriptsize s}}}$~=~(12~$\pm$~3.6)\%.  Taking the data curve in
Fig.~\ref{fig:BsLJCOther}(b) literally, OPAL also excludes
ranges of $\Delta m_{\mbox{\scriptsize s}}$
between 6.3 and 7.9~ps$^{-1}$\/ and above 19.6~ps$^{-1}$ at
95\% Confidence Level.
At 97\%~Confidence Level, however, these exclusions disappear, so their
significance is marginal.

\subsection{Lepton--Lepton Method}
\begin{figure}[t]
\begin{center}
\vspace*{13pt}
\epsfig{file=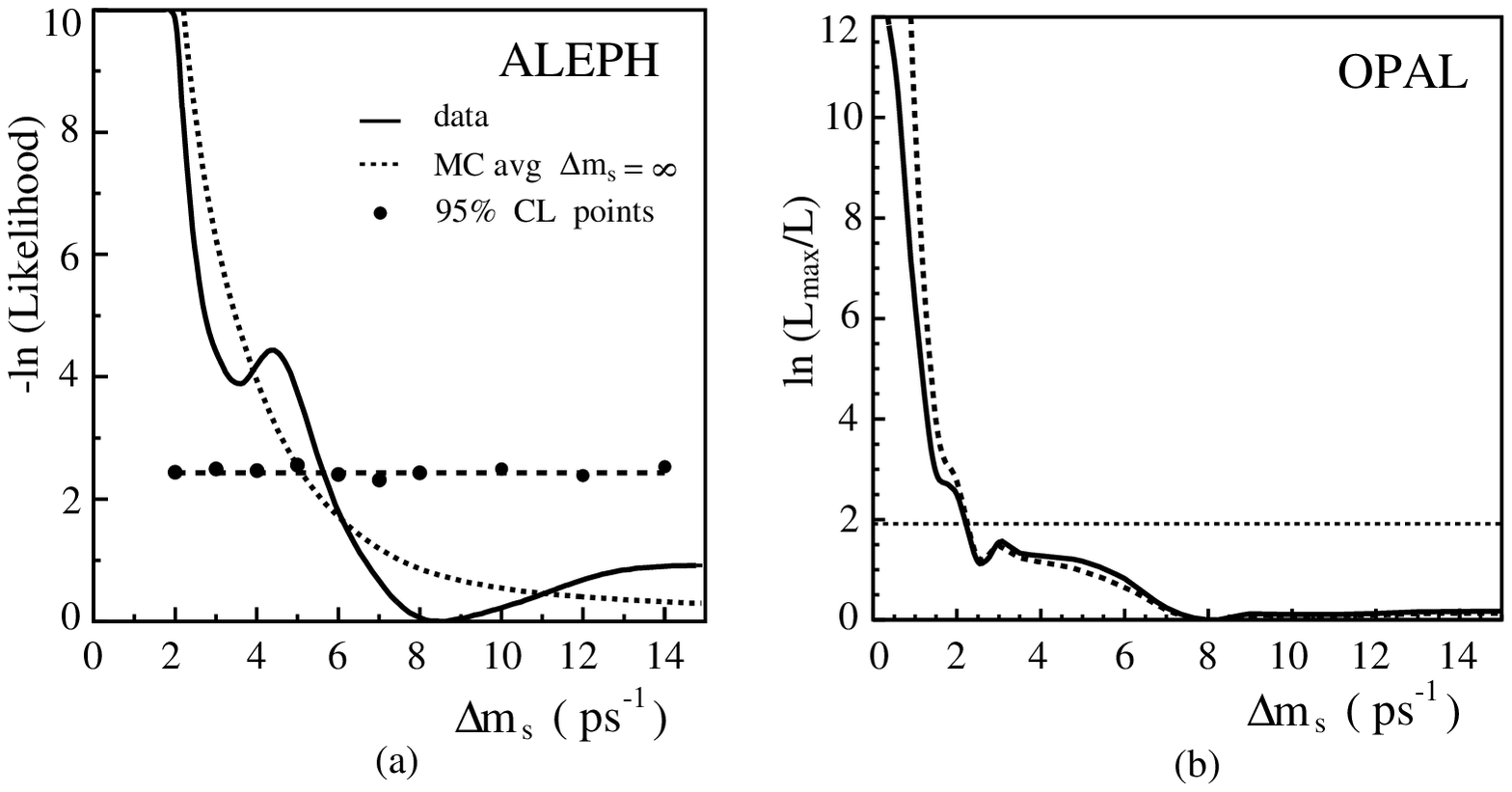,width=6in}
\end{center}
\fcaption{Limit on $\Delta m_{\mbox{\scriptsize s}}$\/ from (a) ALEPH and (b) OPAL
Lepton--Lepton method.}
\label{fig:BsLL}
\end{figure}
This method has been used by the ALEPH\cite{AL:bdall}
and OPAL\cite{OP:bdsll} collaborations
at LEP, and its schematic is that of Fig.~\ref{fig:BdLL}(a) with
the $B_{d}$\/ and
$\bar{B}_{d}$\/ replaced by $B_{\mbox{\scriptsize s}}$\/ and $\bar{B}_{\mbox{\scriptsize s}}$\/.  As with
the $B_{d}$\/ analysis, the $B$\/ decay vertex is formed by the
secondary vertex including a high $p_{t}$\/ lepton, and the decay
and production flavors are tagged by the signs of leptons on the
flight distance and opposite sides.  The results from the ALEPH and
OPAL collaborations are shown in Fig.~\ref{fig:BsLL}.
The preliminary result from
ALEPH\cite{AL:bdall} is $\Delta m_{\mbox{\scriptsize s}}>5.6$~ps$^{-1}$\/ at
95\%~Confidence Level for
$f_{B_{\mbox{\scriptsize s}}}$~=~(12.2~$\pm$~3.2)\% while the published result from
OPAL\cite{OP:bdsll} is $\Delta m_{\mbox{\scriptsize s}}>2.2$~ps$^{-1}$\/ at
95\%~Confidence Level for
$f_{B_{\mbox{\scriptsize s}}}$~=~(12.0~$\pm$~3.6)\%.

\subsection{Lepton--Kaon Correlations Method}
\begin{figure}[t]
\begin{center}
\epsfig{file=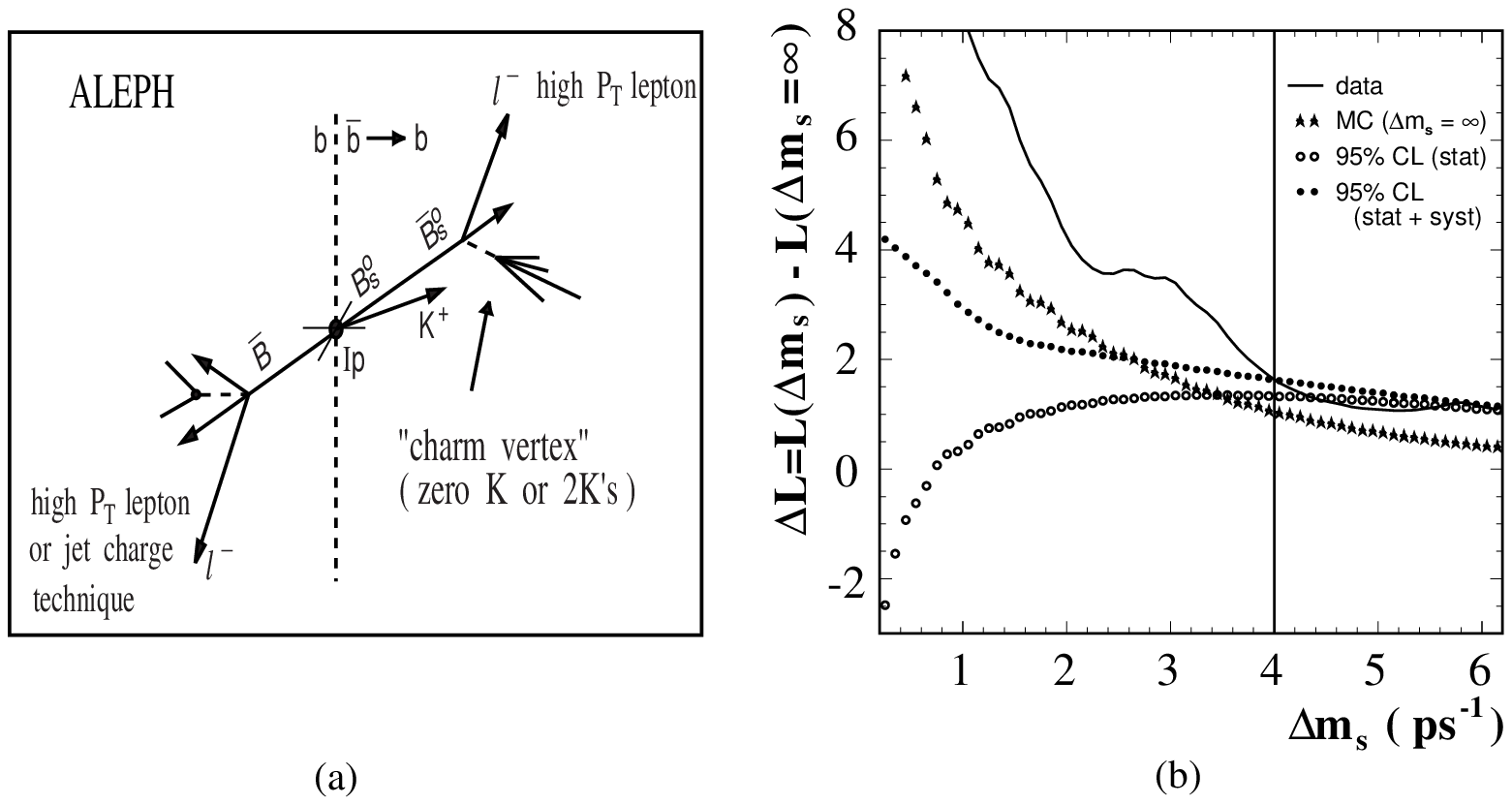,width=6in}
\end{center}
\fcaption{Limit on $\Delta m_{\mbox{\scriptsize s}}$\/ from ALEPH with Lepton--Kaon
Correlations method.}
\label{fig:BsLK}
\end{figure}
This method has been used
by the ALEPH\cite{AL:bslk}
collaboration, and
its schematic is given in Fig.~\ref{fig:BsLK}(a).  In order to
enrich the sample with
$B_{\mbox{\scriptsize s}}$\/ events, the analysis requires that a charged kaon from the primary
vertex be identified.  This kaon must have the opposite sign of the
lepton or jet charge in the opposite hemisphere, in order to improve
the tag of the production flavor.  To enrich the decay
vertex with $D_{\mbox{\scriptsize s}}$\/, the ``charm vertex'' is required to contain
either zero or two kaons, or one kaon with a charge opposite to the lepton.  The decay
flavor is tagged by the sign of the lepton from the decaying $B$\/ meson,
as in other methods.

This selection yields 4436 lepton--kaon correlations, and enriches the
$B_{\mbox{\scriptsize s}}$\/ sample by a factor of 1.35.  The method has a high tag rate
of about 80\%.  The preliminary result for this measurement is shown in
Fig.~\ref{fig:BsLK}(b) giving $\Delta m_{\mbox{\scriptsize s}}>4.0$~ps$^{-1}$ at
95\%~Confidence Level for $f_{B_{\mbox{\scriptsize s}}}$~=~(12~$\pm$~3)\%.

\subsection{Estimation of $B_{\mbox{\scriptsize s}}$\/ fraction in $b$ events.}
There are currently two
methods for determining the fraction of $B_{\mbox{\scriptsize s}}$\/ in an inclusive
lepton sample.  The first method is from $D_{\mbox{\scriptsize s}}\ell$\/ correlations.
ALEPH has measured the product branching ratio\cite{AL:Dslcorr}:
\begin{equation}
f_{B_{\mbox{\scriptsize s}}} \cdot {\rm Br}({\it B}_{\mbox{\scriptsize s}}^{0}\rightarrow
{\it D}_{\mbox{\scriptsize s}}^{-}\ell^{+}\nu X) = 0.82 \pm 0.09 (stat) ^{+ 0.13}_{-0.14} (syst)\%
\label{eq:branch}
\end{equation}
and from this, derives\cite{AL:Dslcorr,AL:Vcb}
$f_{B_{\mbox{\scriptsize s}}}$\/ = 11.0 $\pm$\/ 2.8\%.

The second method uses the average time-integrated mixing parameter
$\bar{\chi}=f_{B_{\mbox{\scriptsize s}}}\chi_{\mbox{\scriptsize s}}+f_{B_{d}}\chi_{d}$\/.  Assuming
$f_{B_{u}}+f_{B_{d}}+f_{B_{\mbox{\scriptsize s}}}+f_{\Lambda_{b}}=1$\/,
$f_{B_{u}}=f_{B_{d}}$, and $\chi_s=0.5$, the $B_{\mbox{\scriptsize s}}$ fraction is given by:
\begin{equation}
f_{B_{\mbox{\scriptsize s}}} = \frac{2 \bar{\chi} - (1-f_{\Lambda_{b}}) \chi_{d}}
                 {1-\chi_{d}}.
\label{eq:fbscalc}
\end{equation}
Using\cite{BJ:Kroll} $\tau_{B_{d}}$\/ = 1.570 $\pm$\/ 0.049 ps,
$\Delta m_{d}$\/ = 0.458 $\pm$\/ 0.020 from the LEP average as given
in Fig.~\ref{fig:dmd}, and the $\Upsilon(4s)$\/
average\cite{Br:Wells} $\chi_{d}(\Upsilon(4s))$ = 0.167 $\pm$\/ 0.025,
the world average of the time-integrated $B_{d}$\/ oscillation parameter,
$\chi_{d}$\/ is calculated to be 0.170~$\pm$~0.011.  Table~\ref{tab:chi}
gives the average
mixing parameter, $\bar{\chi}$~=~ 0.115~$\pm$~0.006.
\begin{table}[tb]
\tcaption{Summary of measurements of $\bar{\chi}$.}
\begin{center}
\begin{tabular}{||c|c||} \hline \hline
Measurement                   &    $\bar{\chi}$                  \\ \hline
LEP+SLD\cite{chi:lep}         &    0.1145 $\pm$ 0.0061           \\
(dileptons)                   &                                  \\ \hline
CDF (e$\mu$)\cite{chi:cdf}    &    0.118 $\pm$ 0.008 $\pm$ 0.020 \\ \hline
CDF ($\mu\mu$)\cite{chi:cdf}  &    0.118 $\pm$ 0.021 $\pm$ 0.026 \\ \hline
D0  ($\mu\mu$)\cite{chi:D0}   &    0.09  $\pm$ 0.04  $\pm$ 0.03  \\ \hline
World Average                 &    0.115 $\pm$ 0.006             \\ \hline
\hline
\end{tabular}
\end{center}
\label{tab:chi}
\end{table}
The baryon fraction, $f_{\Lambda_{b}}$\/, is derived from
$\Lambda_{c}$--lepton and $\Lambda$--lepton correlations, in analogy with
Eq.~(\ref{eq:branch}) above.  Using\cite{RF:PDG}
{\rm Br}($\Lambda_{c}\rightarrow\Lambda X$)~=~35~$\pm$~11\%,
the LEP measurements\cite{chi:lamb} can be averaged to yield
\begin{equation}
f_{\Lambda_{b}}\cdot
Br(\Lambda_{b}\rightarrow\Lambda_{c}X\ell\nu) = 1.67 \pm 0.30 \%,
\end{equation}
with common systematic effects taken into account.  Following
the method given in Ref.~20, the baryon fraction
is then calculated to be
$f_{\Lambda_{b}}$~=~12.8~$\pm$~3.9\%.

With these inputs, $f_{B_{\mbox{\scriptsize s}}}$\/ from the second method
is then 9.9~$\pm$~1.9\%.  An average of the two methods then yields
a final estimate of the fraction of $B_{\mbox{\scriptsize s}}$\/ mesons produced
in $Z\rightarrow b\bar{b}$\/ decay,
$f_{B_{\mbox{\scriptsize s}}}$~=~10.2~$\pm$~1.6\%.

\subsection{Summary of lower limit for $\Delta m_{\mbox{\scriptsize s}}$\/}
Table~\ref{tab:dms} summarizes the lower limits at 95\% Confidence
Level placed on $\Delta m_{\mbox{\scriptsize s}}$\/
from the LEP experiments.
These limits on $\Delta m_{\mbox{\scriptsize s}}$ are computed using different
techniques, and there is currently no combined result which takes
correlated statistical and systematic errors into account.
Thus, the best limits on $B_{\mbox{\scriptsize s}}$\/ oscillation,
$\Delta m_{\mbox{\scriptsize s}} > 6.1$\/ ps$^{-1}$ for $f_{B_{\mbox{\scriptsize s}}}$ = 12$\%$\/ and
$\Delta m_{\mbox{\scriptsize s}} > 5.6$\/ ps$^{-1}$ for $f_{B_{\mbox{\scriptsize s}}}$ = 10$\%$\/ 
using the Lepton--Jet charge method by ALEPH are
taken as the current limits on $\Delta m_{\mbox{\scriptsize s}}$\/.  Defining
$x_{\mbox{\scriptsize s}}=\Delta m_{\mbox{\scriptsize s}}\tau_{B_{\mbox{\scriptsize s}}}$ where\cite{BJ:Kroll}
$\tau_{B_{\mbox{\scriptsize s}}}$~=~1.58~$\pm$~0.10~ps$^{-1}$,
the values of $x_{\mbox{\scriptsize s}}$\/ are shown in Table~3 by shifting
the central
value of $\tau_{B_{\mbox{\scriptsize s}}}$\/ down by $1\sigma$\/.  Using the world average
central values\cite{th:comb} of the quantities in Eq.~(\ref{eq:qcalc})
and including their uncertainties by shifting the values by
$1\sigma$\/ to the conservative side, yields the ratios
$\Delta m_{\mbox{\scriptsize s}}/\Delta m_{d}$\/ and
$|V_{ts}/V_{td}|$\/ as shown in Table~3.

\newpage
\begin{table}[t]
\tcaption{Summary of Limits on $\Delta m_{\mbox{\scriptsize s}}$\/ at 95\% C.L.}
\begin{center}
\begin{tabular}{||l|c|c||} \hline \hline
                        &  $\Delta m_{\mbox{\scriptsize s}}$\/ (ps$^{-1}$\/)  &
                           $f_{B_{\mbox{\scriptsize s}}}$                     \\ \hline
ALEPH (91-94)           &  $> 6.1$                         &
                           12\%                            \\
(lept/Q$_{J}$\/)        &  $> 5.6$                         &
                           10\%                            \\ \hline
ALEPH (91-94)           &  $> 5.6$                         &
                           12$\pm$3\%                      \\
(lept/lept)             &                                  &
                                                           \\ \hline
ALEPH (91-94)           &  $> 4.0$                         &
                           12$\pm$3\%                      \\
(lept/K+$Q_{J}$)                &                                  &
                                                           \\ \hline
DELPHI (91-94)          &  $> 4.2$                         &
                           10$\pm$3\%                      \\
(lept/Q$_{J}$\/)        &                                  &
                                                           \\ \hline
DELPHI                  &  $> 1.5$                         &
                                                           \\
($D_{\mbox{\scriptsize s}}\ell$/Q$_{J}$\/) &                                  &
                                                           \\ \hline
OPAL (91-94)            &  $> 3.3$                         &
                           12.0$\pm$3.6\%                  \\
(lept/Q$_{J}$\/)        &                                  &
                                                           \\ \hline
OPAL (91-93)            &  $> 2.2$                         &
                           12.0$\pm$3.6\%                  \\
(lept/lept)             &                                  &
                                                           \\ \hline \hline
\end{tabular}
\end{center}
\label{tab:dms}
\end{table}
\begin{table}[h]
\label{tab:thyres}
\tcaption{Constraints on physical quantities resulting from measurements
of $\Delta m_{d}$\/ and $\Delta m_{\mbox{\scriptsize s}}$\/.}
\begin{center}
\begin{tabular}{||c|c|c||} \hline \hline
                                       & $f_{B_{\mbox{\scriptsize s}}}=12\%$ &
                                         $f_{B_{\mbox{\scriptsize s}}}=10\%$ \\ \hline
$\Delta m_{\mbox{\scriptsize s}}$                         & $>6.1$ ps$^{-1}$ &
                                         $>5.6$ ps$^{-1}$ \\ \hline
$x_{\mbox{\scriptsize s}}$                                & $>9.0$           &
                                         $>8.3$           \\ \hline
$\Delta m_{\mbox{\scriptsize s}}/\Delta m_{d}$            & $>12.8$          &
                                         $>11.8$          \\ \hline
$\left| V_{ts}/V_{td}\right|$          & $>2.8$           &
                                         $>2.7$           \\ \hline \hline
\end{tabular}
\end{center}
\end{table}

\section{Conclusion}
In summary, by studying the time-dependence of
$B^{0}$\/--$\bar{B}^{0}$\/ oscillations, recent experiments have given

(i) an accurate value for $\Delta m_{d}$\/, the mass difference
between $(B_{d})_{L}$\/ and $(B_{d})_{S}$\/; and

(ii) a lower bound for $\Delta m_{\mbox{\scriptsize s}}$\/, the mass difference between
$(B_{\mbox{\scriptsize s}})_{L}$\/ and $(B_{\mbox{\scriptsize s}})_{S}$\/.

The results from the ALEPH, DELPHI and OPAL Collaborations at LEP
of CERN and the CDF Collaboration at the Tevatron Collider of Fermilab
are summarized in Table~\ref{tab:results}.  In particular,
\begin{equation}
\frac{\Delta m_{d}}
     {\Delta m_{K}} = 85.8 \pm 3.6.
\end{equation}

The impact of the results presented here is shown in
Fig.~15 in the
$(\Delta m_{d},\Delta m_{\mbox{\scriptsize s}})$ plane together with the region allowed
by the Standard Model\cite{th:dmsdmd}.
\begin{table}[p]
\tcaption{Summary of $\Delta m_{d}$\/ and $\Delta m_{\mbox{\scriptsize s}}$\/ results.}
\begin{center}
\begin{tabular}{||l|c|c||} \hline \hline
\multicolumn{3}{||c||} {Mass Differences for the Long and Short Eigenstates}
               \\ \hline
               & $\Delta m$  (ps$^{-1}$\/)     &
$\Delta m$  (eV)             \\ \hline
$\Delta m_{K}$ & $(5.33\pm0.03)\times 10^{-3}$ &
$(3.51\pm0.02)\times 10^{-6}$\\ \hline
$\Delta m_{d}$ &
$0.457 \pm 0.019$&
$(3.01 \pm 0.13)\times 10^{-4}$\\ \hline
$\Delta m_{\mbox{\scriptsize s}}$\/ (95$\%$ C.L.) & & \\
$f_{B_{\mbox{\scriptsize s}}} = 12\%$&
$> 6.1$&
$> 4.0 \times 10^{-3}$ \\
$f_{B_{\mbox{\scriptsize s}}} = 10\%$ &
$> 5.6$ &
$> 3.7 \times 10^{-3}$ \\ \hline \hline
\end{tabular}
\end{center}
\label{tab:results}
\end{table}
\begin{figure}[p]
\label{fig:dmddms}
\begin{center}
\epsfig{file=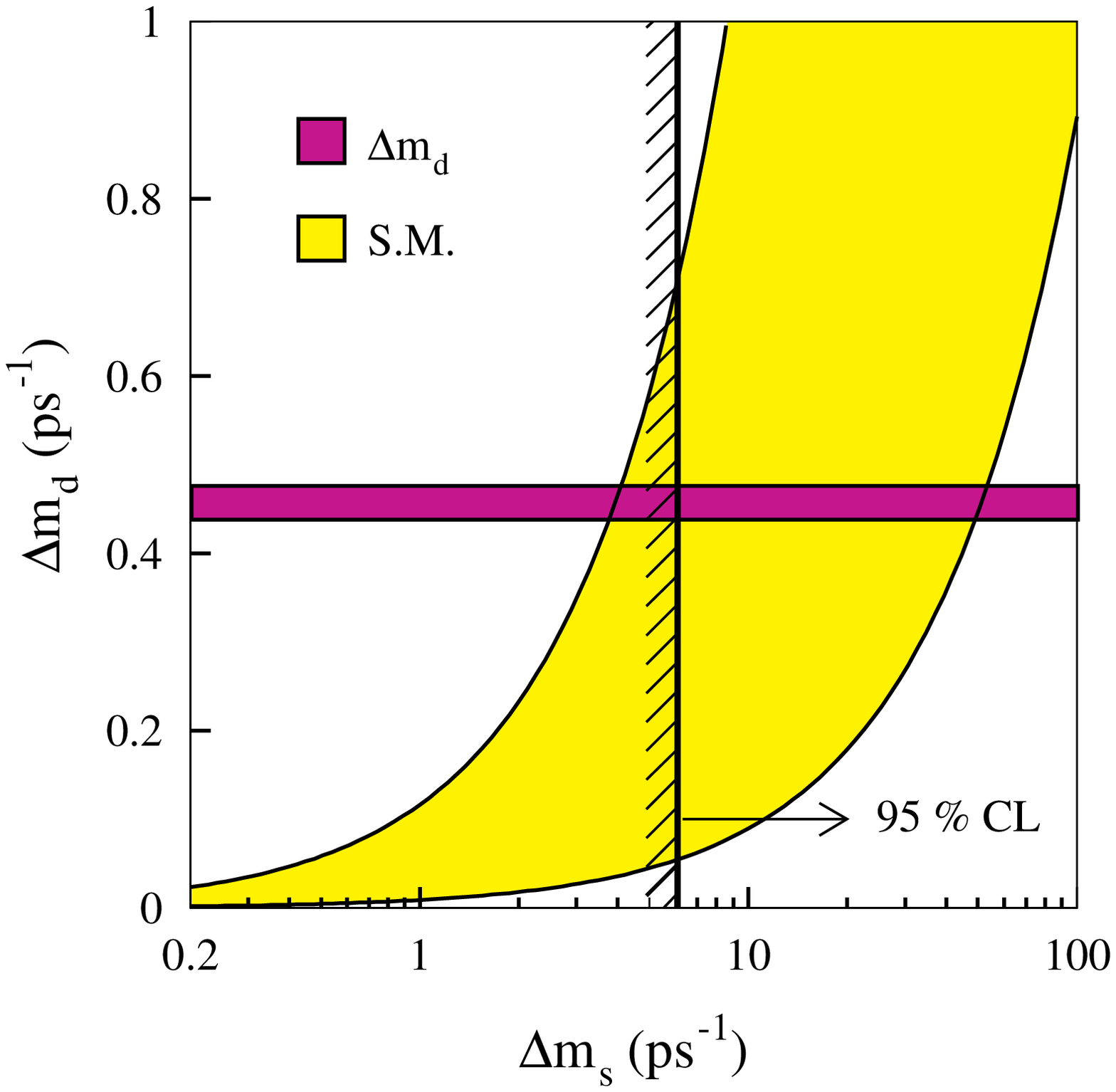,width=6in}
\fcaption{Constraints on the ($\Delta m_{d}$\/, $\Delta m_{\mbox{\scriptsize s}}$\/) plane.}
\end{center}
\end{figure}

\clearpage
\section{Acknowledgement}
I am deeply indebted to Peter McNamara for his invaluable contribution to this talk
and the preparation of this manuscript.  I would like to thank Roger Forty, Sa\'{u}l Gonz\'{a}lez,
Owen Hayes, Hongbo Hu, Hans-Gunther Moser, Yibin Pan, and Min Zheng for their
very valuable discussions.  I am grateful to the United States Department of Energy for
its continuous support through contract DE-AC02-76ER00881 and grant
DE-FG02-95ER40896.

\end{document}

%% file: graphics_sizes.tex
\newdimen\fullfigwidth  

\newdimen\halffigwidth  

\newlength{\fullheight}
\advance \fullheight by -9\baselineskip
\newlength{\halfheight}
\divide \halfheight by 2
\newlength{\thrdheight}
\divide \thrdheight by 3
\newlength{\frthheight}
\divide \frthheight by 4
\newlength{\ffthheight}
\divide \ffthheight by 5
\newlength{\twtdheight}
\divide \twtdheight by 3
\multiply \twtdheight by 2
\newlength{\tweeheight}
\divide   \tweeheight by 3
\multiply \tweeheight by 2
\advance  \tweeheight by 7\baselineskip
\newlength{\twftheight}
\divide   \twftheight by 5
\multiply \twftheight by 2

%% file: eps_graphics.tex
\newread\epsffilein    % file to \read
\newif\ifepsffileok    % continue looking for the bounding box?
\newif\ifepsfbbfound   % success?
\newif\ifepsfverbose   % report what you're making?
\global\newdimen\epsfxsize    % horizontal size after scaling
\global\newdimen\epsfysize    % vertical size after scaling
\newdimen\epsfaxp      % register for aspect ratio test
\global\newdimen\epsftsize    % horizontal size before scaling
\global\newdimen\epsfrsize    % vertical size before scaling
\newdimen\epsftmp      % register for arithmetic manipulation
\newdimen\epsftmx      % register for arithmetic manipulation
\newdimen\epsftmy      % register for arithmetic manipulation
\newdimen\epsfasp      % register for aspect ratio test
\newdimen\epsfxtest      % register for aspect ratio test
\newdimen\pspoints     % conversion factor
\newlength{\epsfxmax}     % maximum horizontal size 

\newlength{\epsfymax}     % maximum vertical size 

\def\epsfbox#1{\global\def\epsfllx{72}\global\def\epsflly{72}%
   \global\def\epsfurx{540}\global\def\epsfury{720}%
   \ifx#1[\let\next=\epsfgetlitbb\else\let\next=\epsfnormal\fi\next{#1}}%
\def\epsfgetlitbb#1#2 #3 #4 #5]#6{\epsfgrab #2 #3 #4 #5 .\\%
   \epsfsetgraph{#6}}%
\def\epsfnormal#1{\epsfgetbb{#1}\epsfsetgraph{#1}}%
\def\epsfgetbb#1{%
\openin\epsffilein=#1
\ifeof\epsffilein\errmessage{I couldn't open #1, will ignore it}\else
   {\epsffileoktrue \chardef\other=12
    \def\do##1{\catcode`##1=\other}\dospecials \catcode`\ =10
    \loop
       \read\epsffilein to \epsffileline
       \ifeof\epsffilein\epsffileokfalse\else
          \expandafter\epsfaux\epsffileline:. \\%
       \fi
   \ifepsffileok\repeat
   \ifepsfbbfound\else
    \ifepsfverbose\message{No bounding box comment in #1; using defaults}\fi\fi
   }\closein\epsffilein\fi}%
\def\epsfsetgraph#1{%
   \epsfrsize=\epsfury\pspoints
   \advance\epsfrsize by-\epsflly\pspoints
   \epsftsize=\epsfurx\pspoints
   \advance\epsftsize by-\epsfllx\pspoints
   \epsfsize{\epsftsize},{\epsfrsize}
   \ifnum\epsfxsize=0 

      \ifnum\epsfysize=0
         \epsfxsize=\epsftsize \epsfysize=\epsfrsize
      \else
         \epsftmp=0pt
         \epsftmp=\epsftsize   \divide\epsftmp\epsfrsize
         \epsfxsize=\epsfysize \multiply\epsfxsize\epsftmp
         \multiply\epsftmp\epsfrsize \advance\epsftsize-\epsftmp
         \epsftmp=\epsfysize
         \loop \advance\epsftsize\epsftsize \divide\epsftmp 2
           \ifnum\epsftmp>0
              \ifnum\epsftsize<\epsfrsize\else
              \advance\epsftsize-\epsfrsize \advance\epsfxsize\epsftmp \fi
         \repeat
      \fi
   \else
      \epsftmp=0pt
      \epsftmp=\epsfrsize   \divide\epsftmp\epsftsize
      \epsfysize=\epsfxsize \multiply\epsfysize\epsftmp   

      \multiply\epsftmp\epsftsize \advance\epsfrsize-\epsftmp
      \epsftmp=\epsfxsize
      \loop \advance\epsfrsize\epsfrsize \divide\epsftmp 2
        \ifnum\epsftmp>0
            \ifnum\epsfrsize<\epsftsize\else
            \advance\epsfrsize-\epsftsize \advance\epsfysize\epsftmp \fi
      \repeat     

   \fi
   \ifepsfverbose\message{#1: width=\the\epsfxsize, height=\the\epsfysize}\fi
   \epsftmp=10\epsfxsize \divide\epsftmp\pspoints
   \centerline{%
   \vbox to\epsfysize{\vfil\hbox to\epsfxsize{%
      \includegraphics{#1}%
                     }                       }%
              }
   \epsfxsize=0pt\epsfysize=0pt
}
{\catcode`\%=12 \global\let\epsfpercent=%\global\def\epsfbblit{%BoundingBox}}%
\long\def\epsfaux#1#2:#3\\{\ifx#1\epsfpercent
   \def\testit{#2}\ifx\testit\epsfbblit
      \epsfgrab #3 . . . \\%
      \epsffileokfalse
      \global\epsfbbfoundtrue
   \fi\else\ifx#1\par\else\epsffileokfalse\fi\fi}%
\def\epsfgrab #1 #2 #3 #4 #5\\{%
   \global\def\epsfllx{#1}\ifx\epsfllx\empty
      \epsfgrab #2 #3 #4 #5 .\\\else
   \global\def\epsflly{#2}%
   \global\def\epsfurx{#3}\global\def\epsfury{#4}\fi}%
\def\epsfsize #1,#2{%
  \epsftmx=0pt \advance\epsftmx by #1  % \showthe\epsftmx
  \epsftmy=0pt \advance\epsftmy by #2  % \showthe\epsftmy
   \ifnum\epsfxmax=0 

        \epsfxsize=0pt
   \else\ifnum\epsfymax=0 

        \epsfxsize=0pt
   \else
     \ifdim\epsftmy>\epsftmx 

        \epsfasp=\epsftmy    \divide\epsfasp\epsftmx
        \epsfxtest=\epsfxmax \multiply\epsfxtest\epsfasp
        \multiply\epsfasp\epsftmx \advance\epsftmy-\epsfasp
        \epsfasp=\epsfxmax
        \loop \advance\epsftmy\epsftmy \divide\epsfasp 2
          \ifnum\epsfasp>0
             \ifnum\epsftmy<\epsftmx\else
               \advance\epsftmy-\epsftmx \advance\epsfxtest\epsfasp \fi
        \repeat
        \ifnum\epsfxtest<\epsfymax
           \epsfxsize=\epsfxmax
           \epsfysize=0pt
        \else
           \epsfxsize=0pt
           \epsfysize=\epsfymax
        \fi
     \else
        \epsfasp=\epsftmx    \divide\epsfasp\epsftmy
        \epsfxtest=\epsfymax \multiply\epsfxtest\epsfasp
        \multiply\epsfasp\epsftmy \advance\epsftmx-\epsfasp
        \epsfasp=\epsfymax
        \loop \advance\epsftmx\epsftmx \divide\epsfasp 2
          \ifnum\epsfasp>0
             \ifnum\epsftmx<\epsftmy\else
               \advance\epsftmx-\epsftmy \advance\epsfxtest\epsfasp \fi
        \repeat
        \ifnum\epsfxtest<\epsfxmax
           \epsfxsize=0pt
           \epsfysize=\epsfymax
        \else
           \epsfxsize=\epsfxmax
           \epsfysize=0pt
        \fi
     \fi
   \fi\fi
   \epsfxmax=0pt \epsfymax=0pt
}
\let\epsffile=\epsfbox
\def\grafeps#1,#2,#3{\vbadness 10000
    \epsfxmax=#2  \epsfymax=#3
    \epsffile{#1}
                    }

%% file: graftwos_macro.tex
\def\graftwos #1,#2,#3,#4,#5{
   \setlength  {\fullfigwidth  } {\columnwidth}
   \setlength  {\halffigwidth  } {\columnwidth}
   \divide\halffigwidth by 2
   \addtolength{\halffigwidth  } {-2ex}
\begin{minipage}[t]{\halffigwidth}
\grafeps {#2},{\halffigwidth},{#1}
{#3}
\end{minipage} \hfill 
\begin{minipage}[t]{\halffigwidth}
\grafeps {#4},{\halffigwidth},{#1}
{#5}
\end{minipage}
}